\documentclass[12pt]{iopart}

\usepackage{graphicx}
\usepackage{epstopdf}
\usepackage{color}
\usepackage{soul}

\begin{document}
\topical[Graphene on hBN]{Graphene on Hexagonal Boron Nitride}
\author{Matthew Yankowitz$^1$, Jiamin Xue$^2$, B. J. LeRoy$^1$}
\address{$^1$ Physics Department, University of Arizona, Tucson, AZ 85721, USA}
\address{$^2$ Microelectronics Research Center, The University of Texas at Austin, 10100 Burnet Rd., Austin, TX 
78758, USA}
\ead{leroy@physics.arizona.edu}

\begin{abstract}
The field of graphene research has developed rapidly since its first isolation by mechanical exfoliation in 2004.  Due to the relativistic Dirac nature of its charge carriers, graphene is both a promising material for next-generation electronic devices and a convenient low-energy testbed for intrinsically high-energy physical phenomena.  Both of these research branches require the facile fabrication of clean graphene devices so as not to obscure its intrinsic physical properties.  Hexagonal boron nitride has emerged as a promising substrate for graphene devices, as it is insulating, atomically flat and provides a clean charge environment for the graphene.  Additionally, the interaction between graphene and boron nitride provides a path for the study of new physical phenomena not present in bare graphene devices.  This review focuses on recent advancements in the study of graphene on hexagonal boron nitride devices from the perspective of scanning tunneling microscopy with highlights of some important results from electrical transport measurements.
\end{abstract}

\maketitle

\tableofcontents

\section{Introduction}\label{Introduction}

Graphene has attracted intense experimental and theoretical efforts since its first isolation and identification in 2004~\cite{Novoselov:2004ub}.  The properties of graphene have been studied experimentally using a wide range of techniques, including electrical transport measurements (for a review, see~\cite{dasSarma:2011br}), optical spectroscopy measurements~\cite{Zhou:2006bt,Ferrari:2007fb}, and local scanning probe measurements~\cite{Morgenstern:2011hh,Deshpande:2012jg,Andrei:2012hy}.  Local scanning probe microscopy plays a unique role among these techniques, providing simultaneous local topographic and electronic information (for example: lattice structure, band structure, and electronic scattering) at the atomic scale. Prior reviews of scanning probe microscopy results have mainly focused on graphene on SiO$_2$~\cite{Morgenstern:2011hh,Deshpande:2012jg,Andrei:2012hy} or graphene on graphite~\cite{Andrei:2012hy} samples.  Since 2010, hexagonal boron nitride (hBN) supported graphene samples have triggered a new surge in graphene research~\cite{Dean:2010jy,Dean:2012ht}.

This review focuses on the electronic properties of graphene on hexagonal boron nitride (hBN) heterostructures accessible with scanning tunneling micrsocopy (STM) and spectroscopy (STS) measurements, as well as the application of these results to transport measurements.  STM and STS measurements show that hBN provides a flatter and cleaner substrate for graphene than SiO$_2$.  As such, hBN permits the observation of intrinsic properties of graphene which were previously obscured by the rough and electronically inhomogeneous SiO$_2$ substrate.  Additionally, the hBN substrate acts as a periodic electric potential for graphene, generating new physical phenomenon which do not exist in suspended or SiO$_2$ supported graphene samples.  In general, these devices are much easier to fabricate and measure than suspended graphene devices while offering nearly comparable performance.  Therefore, many research groups have switched to using hBN substrates for electrical transport measurements of graphene.  These transport experiments benefit especially from the cleaner charge environment leading to improved mobility for both exfoliated~\cite{Dean:2010jy} and chemical vapor deposition grown graphene~\cite{Gannett:2011gs,Kim:2011ju}.  Graphene on hBN heterostructures exhibit effects such as the degeneracy-broken integer quantum Hall effect at low magnetic fields~\cite{Dean:2010jy,Taychatanapat:2011hr,Zomer:2011ft,SanchezYamagishi:2012jv,Young:2012bn,Lee:2014tn}, the fractional quantum Hall effect~\cite{Dean:2011ks}, the quantum spin Hall effect~\cite{Young:2013ft,Amet:2013tk}, an insulating state at the charge neutrality point~\cite{Ponomarenko:2011cj,Hunt:2013ef,Amet:2013gw,Woods:2014co}, Coulomb drag~\cite{Gorbachev:2012bn,Titov:2013gg}, Hofstadter quantization~\cite{Hunt:2013ef,Ponomarenko:2013hl,Dean:2013bv}, magnetic focusing~\cite{Taychatanapat:2013ek}, and many more~\cite{Abanin:2011jq,Wang:2011je,Amet:2012eu,Britnell:2012jp,Britnell:2012dq,Young:2012dv,Masubuchi:2012cz,Haigh:2012dx,Sutar:2012br,Goossens:2012jq,Droscher:2012ir,Zomer:2012gw,Bischoff:2012hf,Campos:2012je,Yu:2013ku,Maher:2013gw,Ponomarenko:2013dr,Britnell:2013ku,Meric:2011tm,Engels:2013fu,Wang:2013ch,Epping:2013kd}.
The presence of the hBN substrate is a critical ingredient in these experiments, not only for increasing the sample cleanliness, but in many cases (the Hofstadter quantization being one particularly nice example) is itself responsible for the observed effect.  

Similarly, hBN can be quite beneficial for the study of graphene-based optical devices as well.  Optical measurements such as spatially resolved Raman spectroscopy can better probe the intrinsic optical response of graphene due to the reduced charge inhomogeneity~\cite{Wang:2012ex,Ahn:2013fa,Forster:2013fi,Chattrakun:2013bt}.  Additionally, they may probe the effects induced by the hBN substrate itself at higher energies normally inaccessible to electrical transport experiments, providing a more complete picture of the physics of these heterostructures over a wide energy range~\cite{Abergel:2013el}.  Unique optical characteristics of graphene on hBN heterostructures may also be utilized to design novel devices, such as re-writable p-n junctions~\cite{Ju:2014dz}.  The combination of local and nonlocal measurements of graphene on hBN heterostructures has resulted in rapid developments in the field, yielding studies of new and interesting physical phenomena and important developments necessary for the production of next-generation electronic devices utilizing these materials.  

We introduce our review in \Sref{Introduction} with a brief discussion of the basic properties of graphene (\Sref{Fundamental Graphene}).  There have already been several comprehensive reviews written about the basic structural and electronic properties of graphene~\cite{dasSarma:2011br,Neto:2009cl}, so we restrict our attention here to the most relevant properties needed to understand the new developments seen in graphene on hBN devices.  We similarly discuss the basic properties of hexagonal boron nitride in~\Sref{Fundamental hBN}.  We cover the basic physics and experimental methods of STM in \Sref{STM}.   We then focus on experimental results in this new research field from the perspective of local scanning probe microscopy.  \Sref{Characterization} discusses basic scanning probe microscopy characterization of graphene on hBN devices and compares them with previous results from graphene on SiO$_2$ devices.  We discuss the results of basic topography measurements in \Sref{Topography} and of basic spectroscopy measurements in \Sref{Spectroscopy}.  In \Sref{SDP}, we discuss the renormalized band structure of graphene on hBN devices due to the periodic electric potential from the hBN substrate.  We discuss the calculation of moir\'e wavelengths seen in graphene on hBN devices in \Sref{Wavelength}.  \Sref{SDP_theory} develops the theory of superlattice Dirac points in graphene on hBN, and \Sref{Exp_SDP} discusses their experimental signatures.  We compare these results to those seen in graphene on metallic crystalline substrates in \Sref{OtherPeriodic}.  In \Sref{Unique} we cover the unique electronic properties which can be seen in graphene on hBN devices due to the cleaner charge landscape.  Sections~\ref{Friedel}, \ref{Manipulation}, and \ref{Magnetic} discuss long-wavelength local density of states oscillations, manipulation of atomic surface adsorbents, and effects seen in an external magnetic field.  In \Sref{Transport} we cover a small sample of important electrical transport measurements which are made feasible only by using an hBN substrate.  Sections~\ref{IQHE} and \ref{FQHE} cover recent transport results demonstrating clean integer and fractional quantum Hall effects.  Sections~\ref{Double} and \ref{Butterfly} discuss results from novel device structures exhibiting features such as an insulating state at charge neutrality and the ``Hofstadter butterfly'' spectrum.  Finally, we conclude our review in \Sref{Conclusion}. 

\subsection{Fundamental graphene properties}\label{Fundamental Graphene}

\subsubsection{Structure and electronic properties}

\begin{figure}[h]
\begin{centering}
\includegraphics[width=3in]{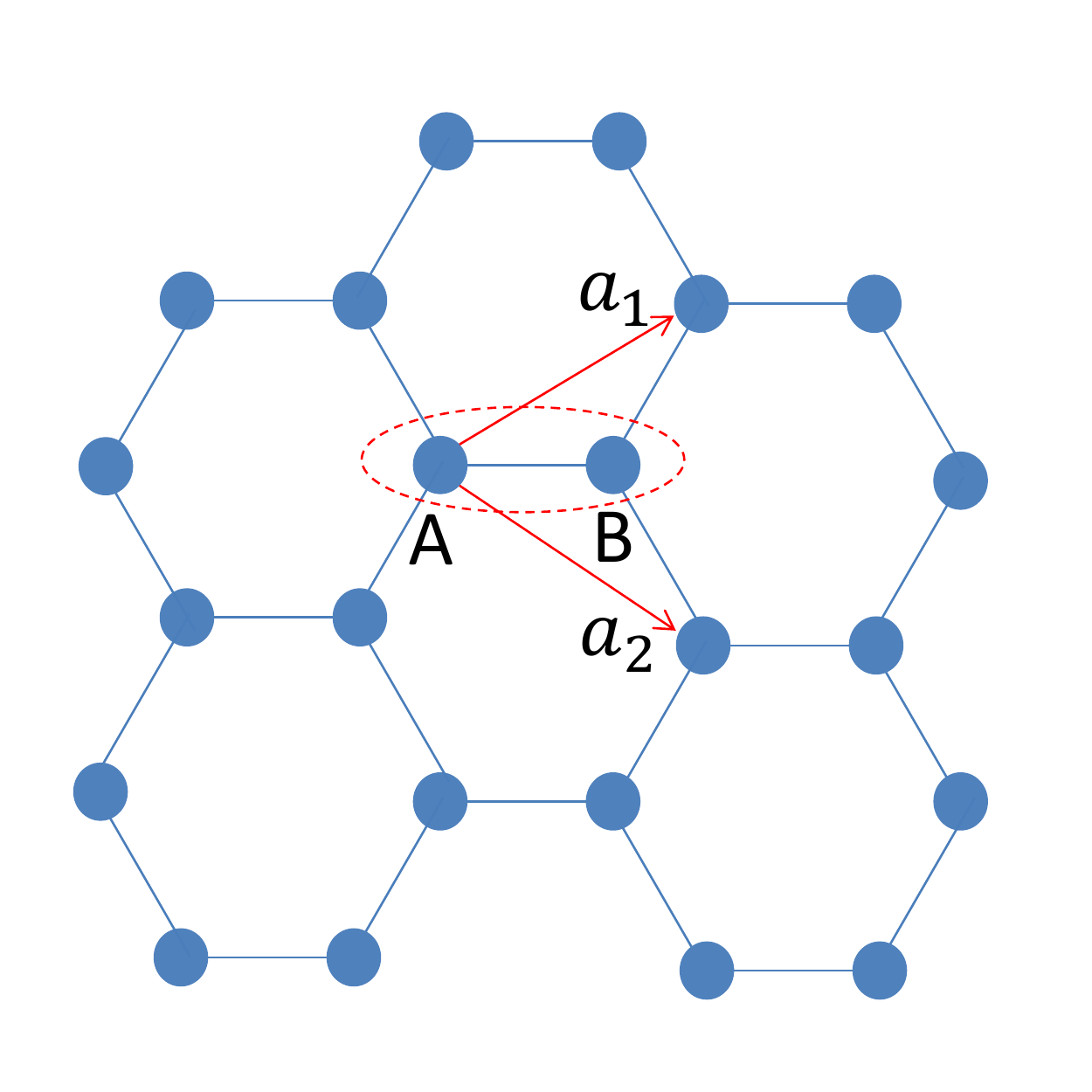}
\caption{Crystal structure of graphene showing the two sublattices labeled A and B.  The two primitive lattice vectors are also shown and labeled as $a_1$ and $a_2$.}
\label{figure0}
\end{centering}
\end{figure}

\begin{figure}[h]
\begin{centering}
\includegraphics[width=3in]{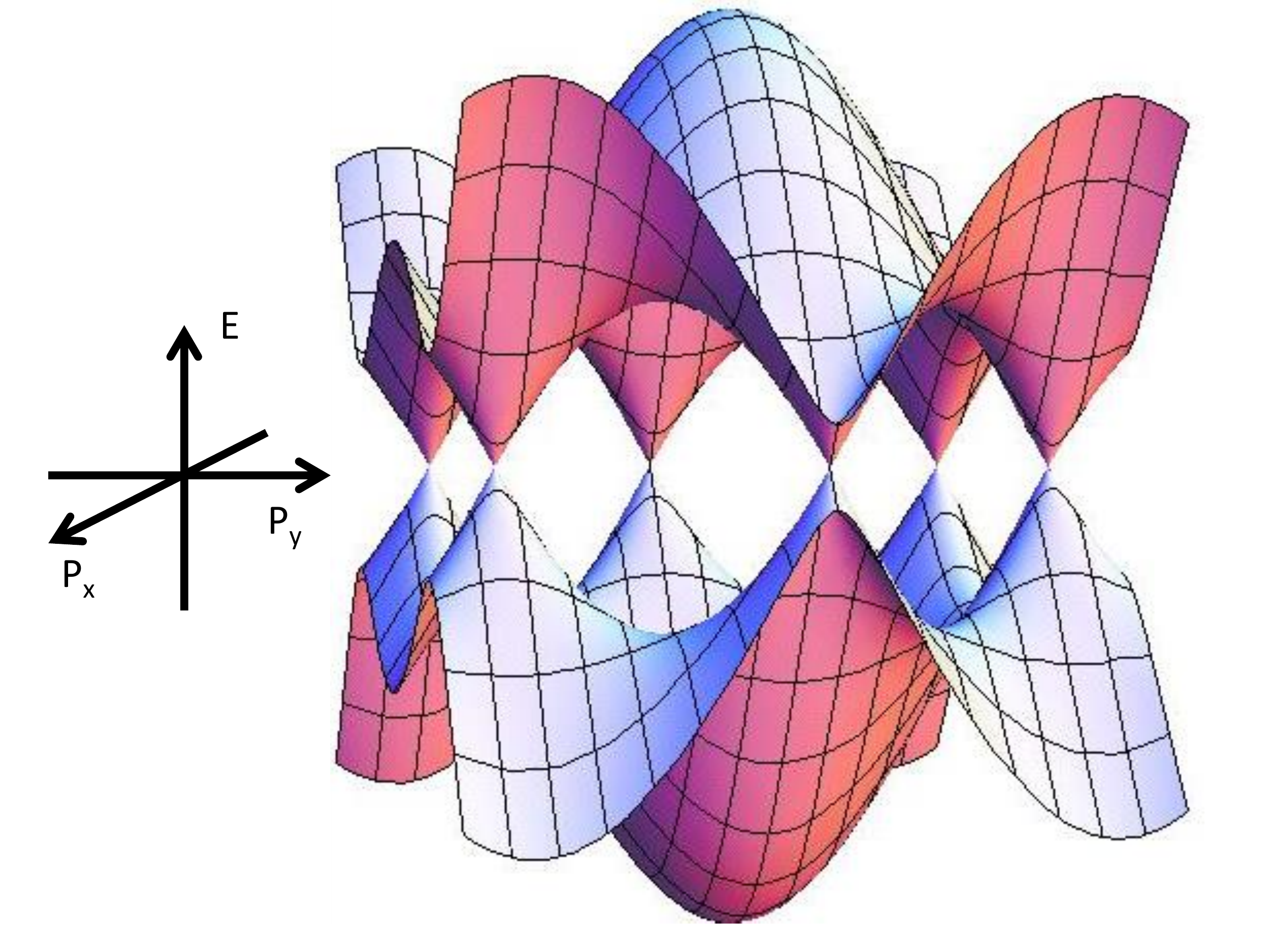}
\caption{Band structure of graphene.  The valence band and conduction band touch at six points at the corners of the first Brillouin zone. Near these points the dispersion relation is approximately linear, resembling that of a massless particle.}
\label{figure1}
\end{centering}
\end{figure}

Graphene is formed by a hexagonal lattice of carbon atoms (figure~\ref{figure0}). It is a Bravais lattice with a two atom basis, conventionally labeled as A and B (figure~\ref{figure0}). The band structure of graphene can be calculated using tight binding theory (see e.g.~\cite{McCann:2012dc}), as shown in figure~\ref{figure1}. The conduction and valence bands touch at 6 points (the so-called Dirac points) at the corners of the first Brillouin zone.  The dispersion relation near these points is approximately linear~\cite{McCann:2012dc}, so the low energy electronic states can be described by a Dirac-like equation:
\begin{equation}
\label{hamiltonian}
H\Psi=
\hbar v_F\left(\begin{array}{cc}
	 0 & k_x-ik_y \\
	 k_x+ik_y & 0
\end{array}\right)\quad
\Psi=E\Psi,
\end{equation}
where $\Psi$ is the wavefunction of an electronic state, $\hbar$ is the reduced Planck constant, $v_F$ is the Fermi velocity, and $k_x$ and $k_y$ are the wave vectors measured from one of the Dirac points.  The eigenvalue and eigenfunction solutions to this equation are
\begin{equation}
\label{eigen}
E_\pm = \pm \hbar v_F k, \quad \quad \psi=\frac{1}{\sqrt{2}}
\left(\begin{array}{c}
  1\\
  \pm e^{i\phi}
\end{array}\right)
e^{i\textbf{k}\cdot\textbf{r}},
\end{equation}
where $k=\sqrt{k_x^2+k_y^2}$, $\phi = \arctan(k_y/k_x)$, and $+$ ($-$) sign corresponds to the conduction (valence) band.  The eigenfunctions are two component vectors and can be characterized by a psuedospin variable, since they formally resemble the real spin vectors of an electron.  The pseudospin originates from the two atoms in the basis of the graphene lattice.  The components of the eigenfunctions represent the relative weight of the wavefunctions of the A- and B-sublattice atoms.  For example, $(1,0)^T$ means the total wavefunction only consists of wavefunctions from A-sublattice atoms. Due to the symmetry between the A- and B-sublattice atoms in graphene, they contribute equal weight to the total wavefunction in Eq.~\ref{eigen}, only differing by a phase factor.  The pseudospin has a deep influence on the electronic properties of graphene, which we will discuss in detail in~\Sref{Friedel}. 

\subsubsection{Behavior in a magnetic field}\label{LL in magnetic field}

In a perpendicular magnetic field, the continuum of electronic states in the graphene band structure breaks up into a series of discrete Landau levels (LLs)~\cite{Neto:2009cl,McCann:2012dc}.  Without considering the zero modes at the Dirac point, the integer quantum hall effect conductance plateaus would lie at $\sigma_{xy}=\pm4Ne^2/h$, where N is an integer, e is the electric charge, h is Planck's constant, and the factor of four represents the two-fold degeneracy in both the spin and valley quantum numbers.  However, at the Dirac point there is a zero mode shared by the electron and hole bands.  Accounting for these two extra states gives the correct expression for the Hall conductivity as $\sigma_{xy}=\pm(N+1/2)4e^2/h$.  Using the linear dispersion relation of graphene $E = \hbar v_F k$, one may calculate the LL energies as  E$_N$ = $sgn(N) \sqrt{2 e \hbar  v_F^2 N B}$, where B is the magnetic field.  Therefore the LL energies follow an unusual square root dependence with magnetic field, as opposed to the usual linear relation in materials with the standard $E \propto k^2$ dispersion. 

\subsubsection{Behavior in a periodic electric potential}\label{periodicpotential}

The chirality of charge carriers in graphene results in the peculiar Klein tunneling behavior~\cite{Katsnelson:2006kd,Stander:2009ce,Young:2009il}, which prevents electrostatic confinement of charge carriers.  Therefore, the possibility of confinement with periodic electric potentials has been explored, both with 1D and 2D potentials~\cite{Park:2008eg,Park:2008kia,Barbier:2008gx,Brey:2009jx,Barbier:2009ks,Sun:2010fx,Burset:2011cc,Ortix:2012tm,Wallbank:2013kz,Wallbank:2013ep,MuchaKruczynski:2013cw,Jung:2013tq,Chen:2014ek}.

For periodic 1D potentials, the group velocity perpendicular to the potential can be strongly reduced, and at certain energies goes to zero~\cite{Park:2008eg,Park:2008kia,Burset:2011cc}.  At these energies, two new Dirac points are created in the graphene band structure (one in the valence band and one in the conduction band, at the same energy relative to the original Dirac point).  These new Dirac points coexist with other states in k-space at the same energy, and therefore the density of states does not go to zero at these new Dirac points for the case of a 1D potential (thus, the original Dirac point remains the only energy with a vanishing DOS).

This also holds for rectangular 2D periodic potentials, as the superlattice Brillouin zone is a rectangle and the chirality of charge carriers in graphene prevents a gap from opening at the new Dirac point energies~\cite{Park:2008eg,Burset:2011cc}.  However, for triangular (or hexagonal) 2D periodic potentials new Dirac points can develop unobstructed by other states in k-space~\cite{Park:2008kia,Ortix:2012tm,Wallbank:2013kz,Wallbank:2013ep,MuchaKruczynski:2013cw,Jung:2013tq,Chen:2014ek}.  The energy at which these new Dirac points develop grows as the superlattice period becomes smaller.

An hBN substrate creates a hexagonal superlattice potential for graphene.  Its effect can be modeled as a linear combination of the bare graphene Hamiltonian (Eq.~\ref{hamiltonian}) and a periodic potential term~\cite{Yankowitz:2012gi}.  It is given by 
\begin{equation}\label{Hgraphene}
\hat{H} = \hbar v_{\rm F} {\bf k} \cdot \vec{\sigma}  + V \sum_\alpha \cos({\bf G}_\alpha
{\bf x})  \, {\cal I} \, ,
\end{equation}
where, ${\bf k}=(k_x,k_y)$, $\vec{\sigma}$ is a vector of Pauli matrices, ${\cal I}$ is the identity matrix, and V is the coupling energy.  The ${\bf G}_\alpha$ are the reciprocal superlattice vectors corresponding to the periodic potential.  We will discuss the specific features of a hexagonal periodic potential created by an hBN substrate in \Sref{SDP_theory}.  

\subsection{Fundamental hexagonal boron nitride properties}\label{Fundamental hBN}

\begin{figure}
\begin{centering}
\includegraphics[width=3in]{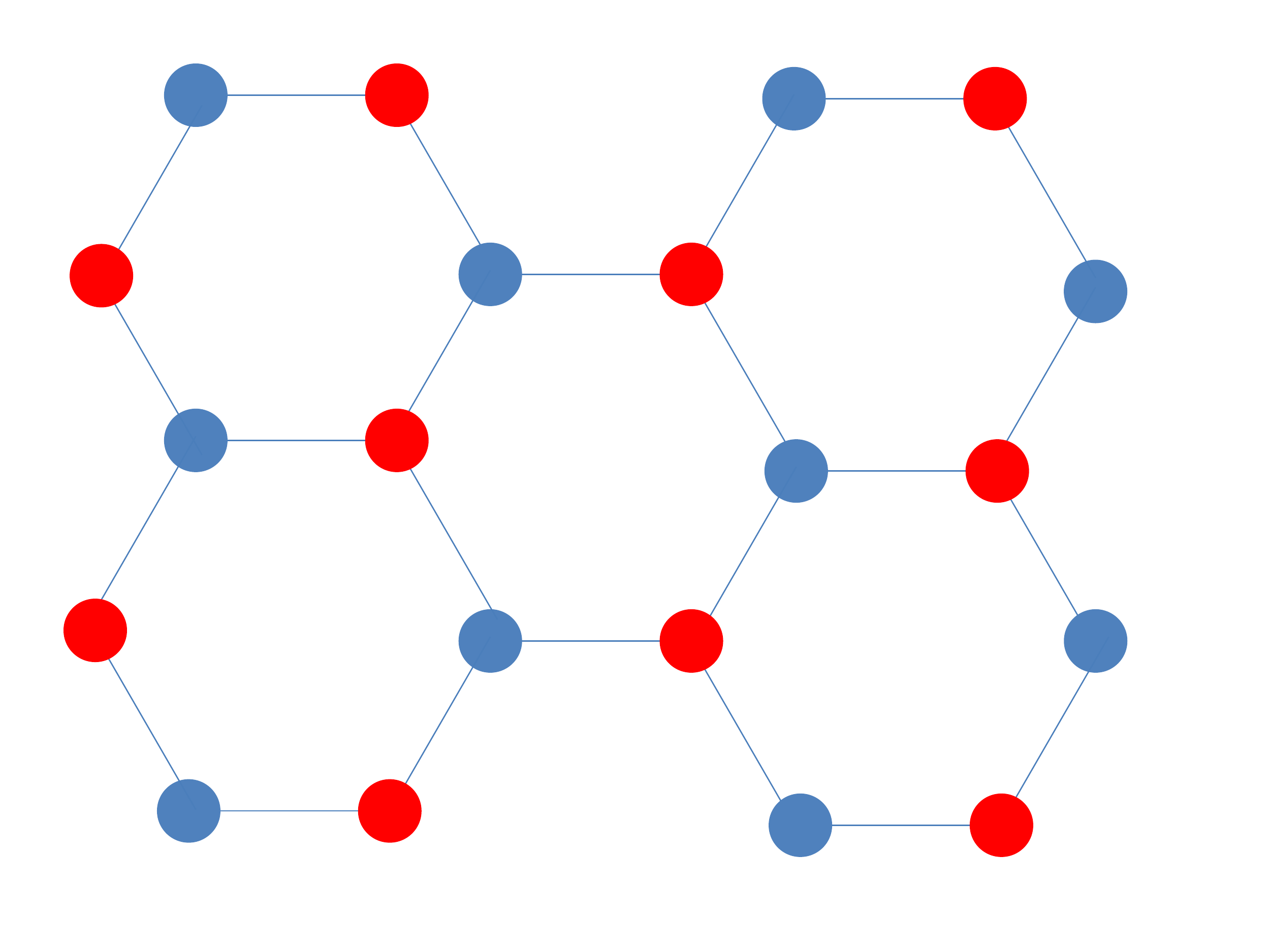}
\caption{Crystal structure of hBN.  The boron atoms are shown in red and the nitrogen atoms are shown in blue.  }
\label{bn}
\end{centering}
\end{figure}

Hexagonal boron nitride is a layered material much like graphite (characterized by strong in-plane bonds and weak van der Waals interaction between layers), which can also be exfoliated to give flakes with atomically flat surfaces.  Each layer of hBN is a hexagonal lattice consisting of boron and nitrogen atoms (figure~\ref{bn}).  Since these two elements neighbor carbon on the periodic table, the lattice mismatch between hBN and graphene is only $\sim$1.8\% (hBN the longer of the two). In graphene, carbon atoms are held together through covalent bonds, but in hBN the boron and nitrogen atoms form ionic bonds. As a result, hBN has a large band gap of $\sim$6 eV~\cite{Watanabe:2004ct}.  These properties make hBN a superior substrate for graphene over amorphously grown SiO$_2$.  Transport measurements of graphene on hBN devices show an improvement of charge carrier mobility by a factor of three to ten compared with the usual graphene on SiO$_2$ devices~\cite{Dean:2010jy,Gannett:2011gs,Kim:2011ju}.  Local measurement techniques are ideal for understanding this improvement and exploring new physical phenomena of this system on the microscopic level.  Scanning tunneling microscopy (STM) and scanning tunneling spectroscopy (STS) are two convenient methods for doing so.

\subsection{Principles of STM}\label{STM}

Scanning tunneling microscopy exploits the quantum tunneling effect to image on the smallest scales achievable by any microscopy technique.  To perform measurements, a sharp metal tip is held very close (typically sub-nanometer range) to the conducting sample of interest.  In this configuration, electrons may tunnel through the vacuum barrier separating the tip and sample, resulting in a small (typically pA-nA range) electrical current between them.  The tunneling current is exponentially sensitive to the separation between the tip and sample, permitting atomic resolution of crystalline samples.  The tunnel current also depends on the density of states (DOS) of both the tip and sample.  To good approximation, the tunneling current can be written as
\begin{equation}
\label{tunnelingCurrent}
I \propto e^{-z \sqrt{8m\Phi/\hbar^2}}\int_0^{eV}\rho_T(E_F-eV+\epsilon)\rho_S(E_F+\epsilon)d\epsilon,
\end{equation}
where $z$ is the tip-sample separation, $m$ is the electron mass, $\Phi$ is the height of the tunnel barrier, $\rho_T$ and $\rho_S$ are DOS of the tip and the sample respectively, $E_F$ is the Fermi energy and $V$ is the bias between them~\cite{Chen:2008vn}.  

\subsubsection{Topography measurements}

To obtain STM topography maps, the bias voltage between the tip and sample is fixed and the tip is scanned across the surface of the sample.  The tunnel current between the tip and sample is held constant by an electrical feedback loop (thus, a constant tunneling resistance is maintained).  The topography of the sample is then inferred by measuring how much the piezo controller must move the tip to maintain the constant tunneling resistance, which is exponentially sensitive to the tip-sample separation.  In addition to fluctuations in sample height, the topography map is convolved with the integrated DOS of both the tip and sample (see Eq.~\ref{tunnelingCurrent}).  Generally the tip DOS is flat, as the tip is a metal, and therefore does not influence the topography measurement.  However, the sample DOS may contain energy dependent features that can influence the topography measurement.  Therefore, all STM topography must be interpreted carefully with this in mind.  

\subsubsection{Spectroscopy measurements}    

Taking a derivative of the tunnel current with respect to the bias voltage yields
\begin{equation}
\label{dIdV}
dI/dV\propto\rho_T(E_F)\rho_S(E_F+eV),
\end{equation}
indicating the differential conductance between the tip and sample is proportional to the product of their DOS.  Again, by choosing a tip with constant DOS, $dI/dV$ yields a measurement proportional to the DOS of the sample.  Experimentally, the differential conductance is measured by turning off the feedback loop, holding the tip at the position and energy of interest, and adding a small AC bias voltage to the DC bias on the tip.  The energy spectrum of interest can be mapped out by varying the DC tip bias as desired.  This measurement technique is known as scanning tunneling spectroscopy.

It is important to note that STS yields a local measure of the density of states of the sample (LDOS), as opposed to transport measurements which probe the global density of states.  LDOS can be defined as
\begin{equation}
\label{LDOS}
\rho (x,E_F+eV)=\mathrm{lim}_{\delta V\rightarrow 0}\frac{1}{e\delta V}\sum _{E_F+eV}^{E_F+eV+\delta V}\left|\psi(x,E)\right|^2,
\end{equation}
where $x$ is the tip position, and $\psi (x, E)$ is the electron's wavefunction. The difference between the DOS and the LDOS is that the latter contains extra information about the electronic wavefunctions.  One of the powerful features of STM is its ability to correlate local topographic and spectroscopic information of the sample. 

\subsubsection{Adatom manipulation}

Additionally, the STM is capable of manipulating adatoms on the surface of a sample and constructing artificial structures consisting of atoms placed at arbitrary positions. The ``quantum corral''~\cite{Crommie:1993vk} of 48 iron adatoms on a copper(111) surface is an elegant demonstration of this method.  By manipulating adatoms, different structures such as dimers, trimers and molecules can be created.  For standard topography and spectroscopy measurements, the STM tip is kept far enough from the surface (controlled via the tunneling resistance) so as not to disturb the mobile surface adatoms.  Adatoms may be manipulated along the surface by reducing the tip-adatom separation until a sufficient wavefunction overlap occurs, binding the adatom to the STM tip.  Then the adatom is moved to the desired location and the tunneling resistance is quickly increased to place the adatom.  This process may be repeated for multiple adatoms to create arbitrary configurations on a surface.  

In Sections~\ref{Characterization}--\ref{Unique} we will discuss how these three modes of STM operation are utilized to study the properties of graphene on hBN systems.

\section{Characterization of graphene on hBN devices}\label{Characterization}

\subsection{Topography}\label{Topography}

\begin{figure}
\begin{centering}
\includegraphics[width=3in]{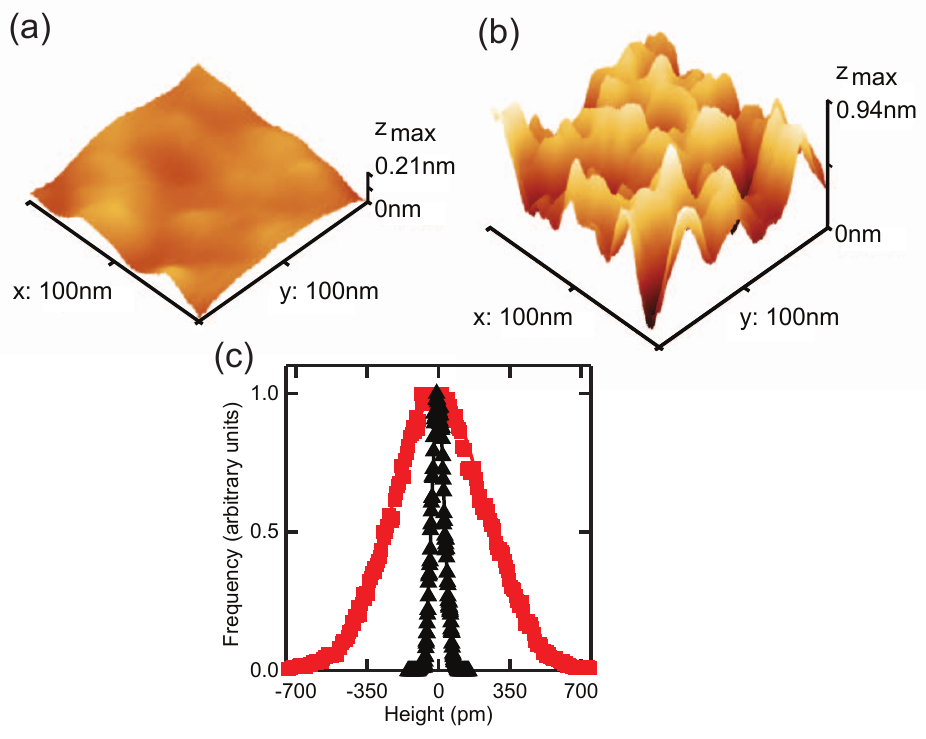}
\caption{(a) STM topography of graphene on hBN. (b) STM topography of graphene on SiO$_2$.  (c) Histogram of roughness of graphene on hBN (black) and graphene on SiO$_2$ (red).  Graphene on hBN exhibits $rms$ roughness nearly an order of magnitude smaller than graphene on SiO$_2$. Adapted with permission from Macmillian Publishers Ltd.: Xue et al., Nature Materials \textbf{10}, 282 (2011), copyright 2011.}
\label{roughness}
\end{centering}
\end{figure}

One of the drawbacks of graphene on silicon oxide devices is that the oxide is thermally grown and is therefore amorphous.  This leads to an $rms$ roughness on the order of 0.5 nm for SiO$_2$ surfaces.  Since graphene tends to conform to its substrate~\cite{Cullen:2010ib}, graphene on SiO$_2$ tends to exhibit a surface roughness of the same magnitude.  However, graphene on hBN is expected to have very low surface roughness due to the atomically flat nature of the crystalline hBN surface.  Experimentally, this can be clearly seen by comparing STM topography maps of graphene on hBN (figure~\ref{roughness}(a)) and graphene on SiO$_2$ (figure~\ref{roughness}(b)).

The histograms of these two images are shown in figure~\ref{roughness}(c).  Both curves are well fit by Gaussian distributions with standard deviations of 224.5 $\pm$ 0.9 pm for graphene on SiO$_2$ and 30.2 $\pm$ 0.2 pm for graphene on hBN. The surface roughness for graphene on hBN is similar to graphene on HOPG~\cite{Lui:2009kv}, suggesting it has reached its ultimate limit of flatness.  This increased flatness helps expose electronic effects in graphene which would be obscured in rougher samples.  While most of the effects discussed in this review benefit from increased graphene flattness, the most striking example of this is the local density of states oscillations from a step edge discussed in \Sref{Friedel}.

\subsubsection{Topographic moir\'e patterns}

\begin{figure}
\begin{centering}
\includegraphics[width=3in]{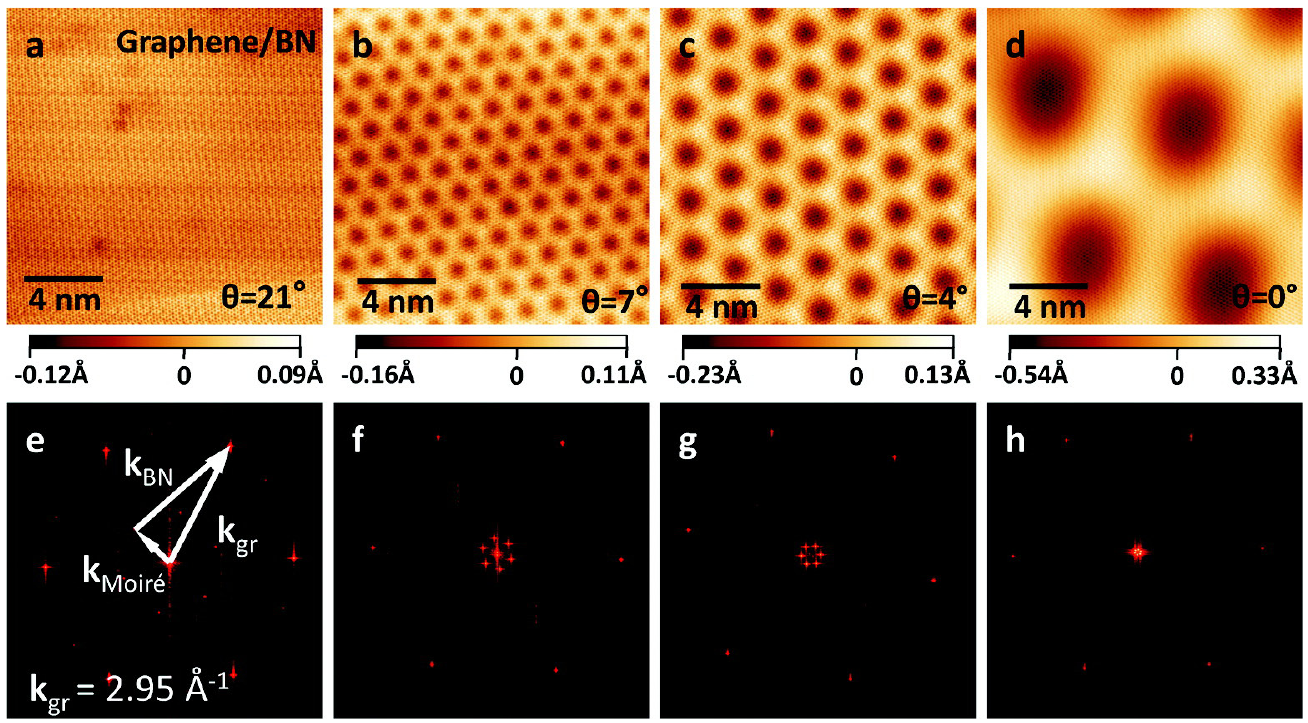}
\caption{(a) - (d) STM topography images of graphene on hBN with different relative lattice rotations (indicated on the image) exhibiting hexagonal superlattices. (e) - (h) Fourier transforms of (a) - (d) showing hexagonal moir\'e spots near the center of the image and atomic lattice points near the outer edges of the image. Reprinted with permission from Decker et al., Nano Lett. \textbf{6}, 11 (2011). Copyright 2011 American Chemical Society.}
\label{fig:SuperlatticeFFT}
\end{centering}
\end{figure}

\begin{figure}
\begin{centering}
\includegraphics[width=4in]{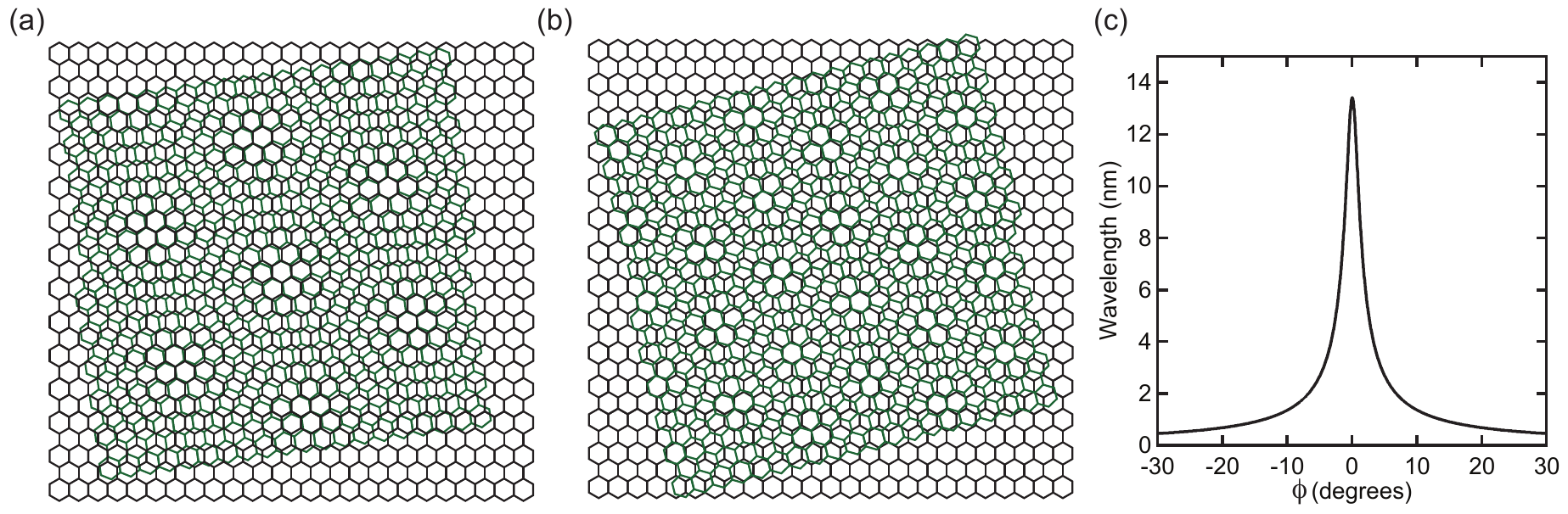}
\caption{Moir\'e pattern schematic for graphene on hBN with different relative rotations.  (a) The rotation is 8 degrees between the lattices.  (b) The rotation is 14 degrees between the lattices giving a shorter moir\'e pattern. (c) Wavelength of the moir\'e pattern for graphene on hBN.}
\label{moire2}
\end{centering}
\end{figure}

Interesting superstructures are observed when examining atomically resolved topography images of graphene on hBN (figures~\ref{fig:SuperlatticeFFT}(a) - (d))~\cite{Yankowitz:2012gi,Xue:2011dv,Decker:2011wh,Roth:2013ca,Yang:2013ev,Tang:2013hy}.  Longer wavelength hexagonal topography modulations are present in addition to the hexagonal structure of the graphene atomic lattice. This hexagonal superstructure, often referred to as a moir\'e pattern, occurs when two lattices with different lengths and/or orientations are placed on top of each other.  Moir\'e patterns often form when graphene is grown on metal substrates~\cite{Wintterlin:2009bc}.  Moir\'e patterns in graphene have also been observed when graphene is placed on other crystalline substrates such as HOPG~\cite{Li:2009eh}, SiC~\cite{Rutter:2007epa}, Cu~\cite{Gao:2010iz}, Ir~\cite{Ndiaye:2006cr}, Ni~\cite{Dedkov:2010jh}, Pd~\cite{Kwon:2009kn}, Pt~\cite{Sutter:2009ff}, Rh~\cite{Roth:2011ja}, Ru~\cite{Marchini:2007gt}.  The wavelengths of the two lattices, as well as their relative rotation, determines the wavelength of the moir\'e pattern.  As shown in figures~\ref{moire2}(a) and (b), smaller rotation angles give longer period moir\'e wavelengths.  For hexagonal crystals with identical lattice constants (such as twisted bilayer graphene), the moir\'e wavelength goes to infinity as the rotation angle approaches zero (perfect alignment).  Since the lattice constant of hBN is about 1.8\% longer than that of graphene, even perfect crystalline alignment of graphene on hBN exhibits a maximum moir\'e wavelength of about 14 nm~\cite{Yankowitz:2012gi}.

\subsection{Spectroscopy}\label{Spectroscopy}

\subsubsection{Density of states}

Early theoretical calculations predicted that a band gap of $\sim$50 meV is opened in graphene on hBN~\cite{Giovannetti:2007jc,Siawinska:2010kr}.  This calculation was based on a lowest-energy lattice configuration where the lattice mismatch is neglected and zero relative rotation is assumed.  In this configuration, one sublattice of carbon atoms sits above boron atoms and the other sits above the center of a hexagon.  This breaks the sublattice symmetry in graphene, and a band gap opens as a result.  However, scanning probe measurements of graphene on hBN devices thus far have always shown a moir\'e pattern of no longer than 14 nm.  These results are from both transferring of graphene on hBN flakes~\cite{Yankowitz:2012gi,Xue:2011dv,Decker:2011wh} and direct chemical vapor deposition growth of graphene on hBN~\cite{Roth:2013ca,Yang:2013ev,Tang:2013hy}.  These results suggest that the change in lattice constant needed for perfect stacking of graphene and hBN is not possible.  Consequently, while at a particular site the A sublattice of graphene may sit above a boron atom, it will reside above a nitrogen within one period of the moir\'e pattern.  Averaging over the entire sample restores sublattice symmetry, and is therefore not expected to open a sizable band gap~\cite{Sachs:2011jp,Kindermann:2012jz}.

Ref~\cite{Xue:2011dv} tests this prediction directly with STS.  Figure~\ref{puddles}(a) plots the dI/dV spectroscopy (which is proportional to the LDOS) of graphene on hBN as a function of STM tip bias.  The dI/dV spectroscopy has a minimum near zero tip bias and increases nearly linearly with energy.  This is in agreement with the linear density of states expected from the band structure of pristine graphene.  This measurement does not exhibit a region of zero dI/dV at the charge neutrality point anticipated if there were a band gap opened due to the hBN substrate, further suggesting the absence of a sizable gap.  Recent transport experiments have suggested the possibility of local lattice commensurability for very long moir\'e wavelengths, which could open spatially dependent band gaps~\cite{Hunt:2013ef,Woods:2014co}.  To date, STM work has been unable to conclusively address this issue.

\subsubsection{Charge variation}

\begin{figure}
\begin{centering}
\includegraphics[width=3in]{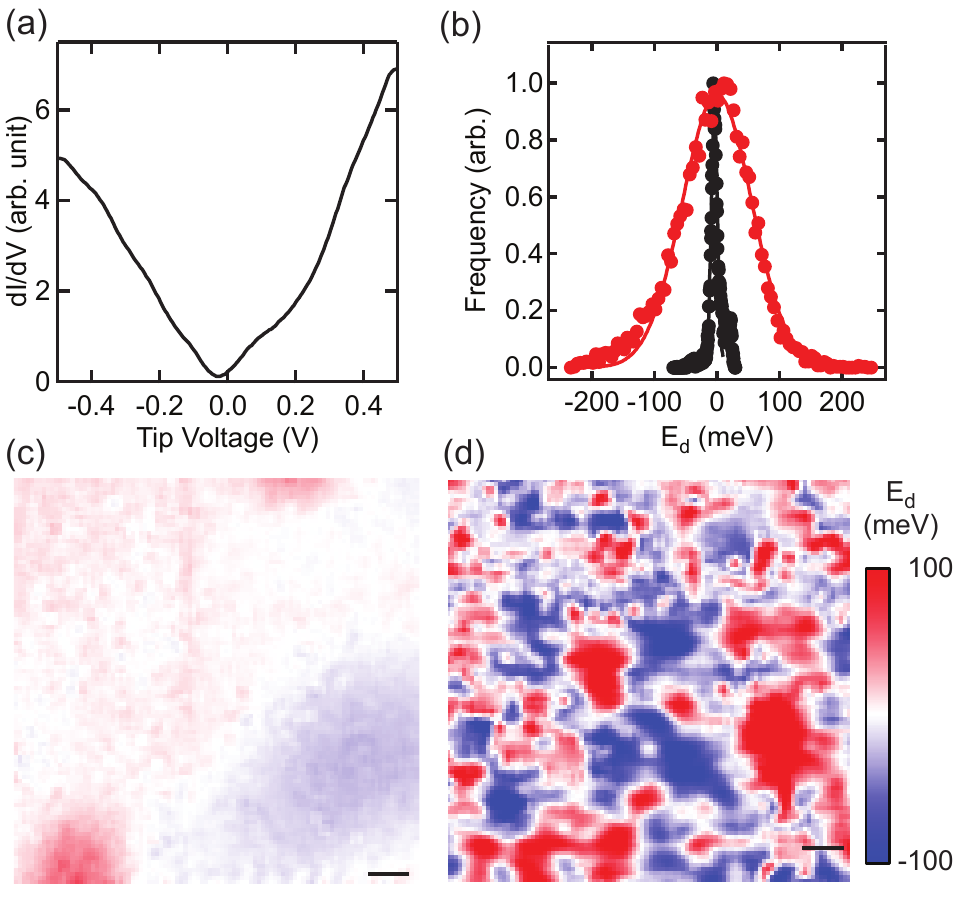}
\caption{(a) dI/dV spectroscopy of graphene on hBN. (b) Histogram of the Dirac point energies with Gaussian fits from (c) and (d) in black and red, respectively. (c) Tip voltage of the Dirac point as a function of position for graphene on hBN. (d) Tip voltage of the Dirac point as a function of position for graphene on SiO$_2$. Scale bar is 10 nm in both. Adapted with permission from Macmillian Publishers Ltd.: Xue et al., Nature Materials \textbf{10}, 282 (2011), copyright 2011.}
\label{puddles}
\end{centering}
\end{figure}

Another drawback to graphene on SiO$_2$ devices is that the substrate causes substantial charge inhomogeneity in the graphene.  This manifests as electron and hole doped regions when the Fermi energy is tuned near the Dirac point.  Scanning probe studies have mapped these charge puddles on SiO$_2$ devices and found that they have a typical size scale on the order of 10 nm and a charge variation on the order of 10$^{11}$ cm$^{-2}$~\cite{Martin:2008ca,Deshpande:2009hx,Zhang:2009ce}.  Transport measurements~\cite{Dean:2010jy,Gannett:2011gs,Kim:2011ju} show that graphene on hBN devices exhibit greatly enhanced charge carrier mobility compared to graphene on SiO$_2$ devices, indicative of reduced charge inhomogeneity.  Scanning probe microscopy measurements of graphene on hBN samples permits direct visualization of this improvement~\cite{Xue:2011dv,Decker:2011wh,Burson:2013fc}.  Charge fluctuations resulting from impurities in the underlying substrate manifest themselves as local shifts in the energy of the Dirac point.  Thus, these impurities act as local dopants for the graphene.  Figures~\ref{puddles}(c) and (d) show spatial maps of the energy of the Dirac point for graphene on hBN and SiO$_2$, respectively.  The blue (red) areas have excess electrons (holes) compared with neutral graphene, so their Dirac points are shifted to more negative (positive) energies.  

The graphene on hBN sample exhibits much smaller shifts in the Dirac point and hence much less charge variation than the graphene on SiO$_2$ sample.  Histograms of these two maps (figure~\ref{puddles}(b)) show that the FWHM value of the Dirac point shift in graphene on hBN is $5.4\pm 0.2$ meV, while for graphene on SiO$_2$ it is $55.6\pm 0.7$ meV. These fluctuations in energy can be converted to the fluctuation in charge carrier densities using the relation $n\propto E^2$ for graphene.  The charge density fluctuations are reduced by about two orders of magnitude when using hBN as a substrate rather than SiO$_2$.  Suspended graphene samples generally have even lower charge fluctuation, however measuring such samples via STM is challenging as the charged STM tip attracts the graphene sheet.  Substrate supported measurements, such as graphene on hBN, are much more stable and are therefore more powerful probes of the intrinsic graphene physics. 

\section{Superlattice Dirac points}\label{SDP}

\subsection{Calculation of moir\'e wavelength}\label{Wavelength}

The hBN substrate creates a superlattice potential that has a profound influence on the electronic properties of graphene.  In general, given a fractional lattice mismatch of $\delta$ between graphene and hBN and a relative rotation angle $\phi$ between the two lattices, the moir\'e wavelength $\lambda$ is uniquely determined by
\begin{equation}\label{eq:lambda}
\lambda = \frac{(1+\delta)a}{\sqrt{2(1+\delta)(1-\cos(\phi))+\delta^2}} \, ,
\end{equation}
where $a$ is the graphene lattice constant.  The relative rotation angle $\theta$ of the moir\'e pattern with respect to the graphene lattice is given by
\begin{equation}\label{eq:theta}
\tan(\theta) = \frac{\sin(\phi)}{(1+\delta)-\cos(\phi)} \, .
\end{equation}
Figure~\ref{moire2}(c) plots the superlattice wavelength as a function of the angle $\phi$ between the graphene and hBN lattices.  The moir\'e wavelength is only long for a small range of rotation angles close to zero degrees.  There is also a maximum wavelength of about 14 nm given by the lattice mismatch $\delta$~\cite{Yankowitz:2012gi}.

STM and conductive AFM measurements are able to clearly detect this moir\'e pattern~\cite{Hunt:2013ef,Ponomarenko:2013hl,Dean:2013bv,Yankowitz:2012gi,Xue:2011dv,Decker:2011wh,Roth:2013ca,Yang:2013ev,Tang:2013hy}.   Figures~\ref{fig:SuperlatticeFFT}(e) - (h) show Fourier transforms of the topography measurements shown in figures~\ref{fig:SuperlatticeFFT}(a) - (d), where both the graphene lattice and superlattice are visible.  The rotation between these two lattices matches that predicted by Eq.~\ref{eq:theta}.  These local scanning probe measurements can be used to quickly and accurately characterize the relative rotation angle between the graphene and hBN lattices.  Raman spectroscopy may also be used to identify long wavelength moir\'e patterns, as the Raman spectra of graphene is influenced by the interaction with the hBN substrate~\cite{Eckmann:2013da}.

\subsection{Theoretical predictions for graphene on hBN}\label{SDP_theory}

\begin{figure}
\begin{centering}
\includegraphics[width=3in]{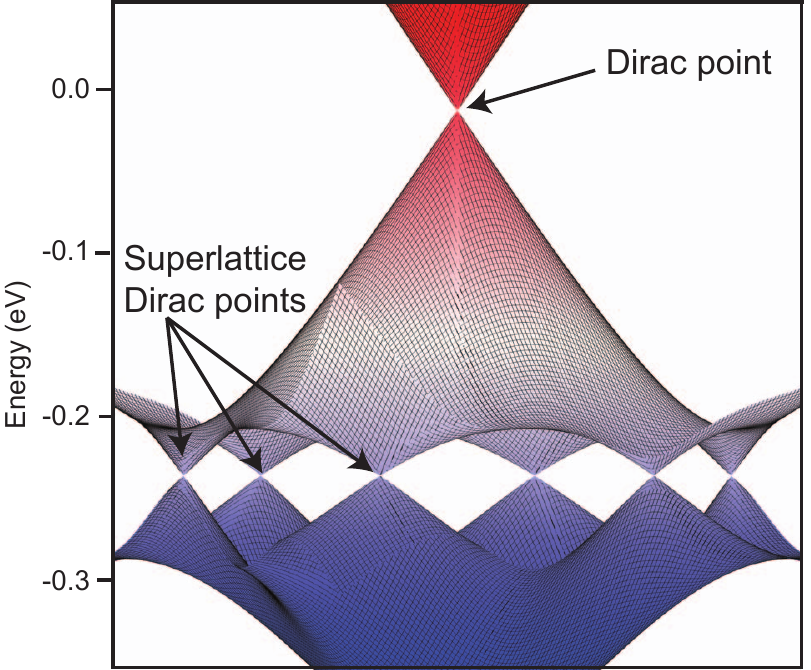}
\caption{Band structure of graphene on hBN.  In addition to the original Dirac point, there are six superlattice Dirac points shown in the valence band.}
\label{fig:SuperlatticeBands}
\end{centering}
\end{figure}

We discussed a model Hamiltonian for graphene on hBN in \Sref{periodicpotential}.  The electronic effect of an hBN substrate on the local density of states is modeled in Ref.~\cite{Yankowitz:2012gi} by considering the interlayer hopping between the graphene and hBN layers.  The potential strength V is estimated to be 0.06 eV from second order perturbation theory.  The reciprocal superlattice vector ${\bf G}_1 = (4 \pi/\sqrt{3} \lambda) (\cos \theta, \sin \theta)$ is determined by the relative rotation of the graphene and hBN lattices according to Eqs.~(\ref{eq:lambda}) and (\ref{eq:theta}).  The two other superlattice wave vectors are obtained by two rotations of 60$^{\circ}$.  The superlattice potential permits ${\bf k} \rightarrow -{\bf k}$ backscattering processes along the direction of the reciprocal lattice vectors ${\bf G}_\alpha$ which are normally forbidden in bare graphene due to chirality.  As long as the potential maintains the sublattice symmetry of the graphene layer, the chirality of the Dirac fermions prevents the opening of a bandgap as is typical for such a process with Schr\"odinger fermions.  Instead, a new set of Dirac points is opened in the valence and conduction bands at the energy where the periodic potential connects the ${\bf k}$ and -${\bf k}$ bands (that is, when 2${\bf k}$ = ${\bf G}_\alpha$).  Assuming the linear band structure of graphene, we expect this energy to be $E = \hbar v_{\rm F}|{\bf G}|/2 = 2\pi \hbar v_{\rm F} /\sqrt{3}\lambda$.  Figure~\ref{fig:SuperlatticeBands} shows a tight-binding model calculation of the band structure of graphene on hBN, where the new set of six valence band Dirac points are visible in addition to the original Dirac point.  There are also six new Dirac points created in the conduction band.

\begin{figure}
\begin{centering}
\includegraphics[width=3in]{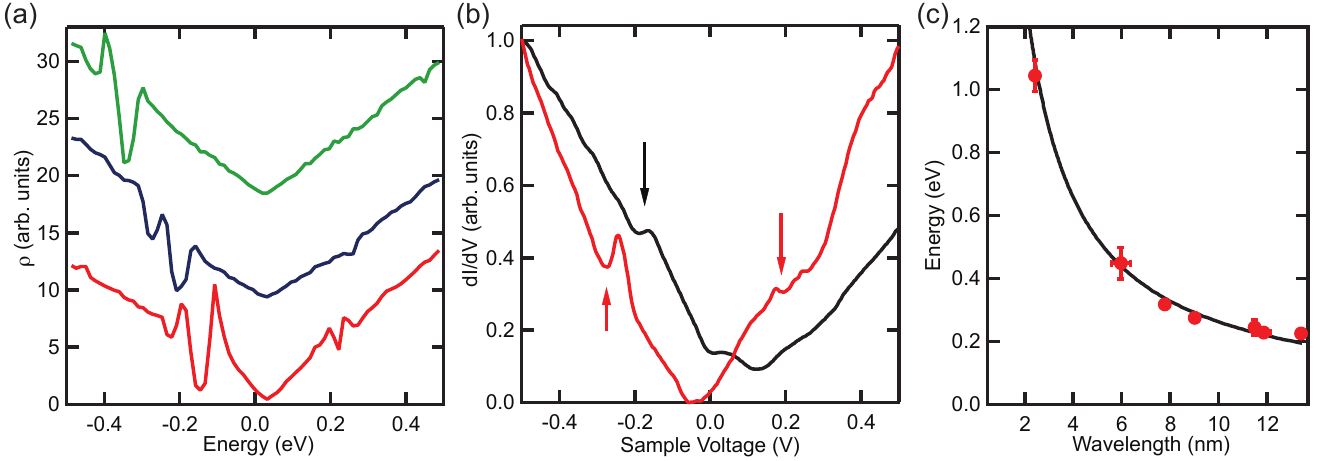}
\caption{(a) Theoretical LDOS for graphene on hBN.  Red is $\phi=0.5^o$ (12.5 nm), blue is $\phi=1^o$ (10.0 nm) and green is $\phi=2^o$ (6.3 nm). (b) Experimental dI/dV spectroscopy for 9.0 nm (black) and 13.4 nm (red) moir\'e wavelengths. (c) Energy of the superlattice Dirac points as a function of wavelength.  Red points are measured experimentally, black curve is the theoretical dependence. Adapted with permission from Macmillian Publishers Ltd.: Yankowitz et al., Nature Physics \textbf{8}, 382 (2012), copyright 2012.}
\label{fig:SDPEnergy}
\end{centering}
\end{figure}

Numerical calculations of the LDOS in the graphene layer provide further evidence for these new Dirac points.  As shown in figure~\ref{fig:SDPEnergy}(a) for a variety of moir\'e wavelengths, the LDOS exhibits two dips (superlattice Dirac points) symmetrically placed about the original Dirac point.  The dip in the LDOS in the valence band is much stronger than the one in conduction band due to the inclusion of next-nearest neighbor interlayer coupling terms, which breaks electron-hole symmetry by inducing modulated hopping between different graphene sublattices~\cite{Yankowitz:2012gi}.  As expected, the superlattice Dirac points move to higher energies with increasing rotation between the graphene and hBN lattices (shorter wavelength moir\'e patterns).

\subsection{Experimental signatures}\label{Exp_SDP}

STS measurements exhibit similar dips in the LDOS symmetrically placed about the original Dirac point (figure~\ref{fig:SDPEnergy}(b))~\cite{Yankowitz:2012gi}, providing evidence for the existence of the superlattice Dirac points.  The same electron-hole asymmetry is observed in these measurements as predicted by the numerical calculations.  In Ref.~\cite{Yankowitz:2012gi}, seven different rotation angles of graphene on hBN were measured.  Only moir\'e wavelengths longer than 2 nm were measured due to the smeared spectroscopy resolution for energies outside the range of $\sim\pm$1 V.  Figure~\ref{fig:SDPEnergy}(c) plots the measured energies of the superlattice Dirac points in these samples as a function of moir\'e wavelength.  The black curve plots the expected dispersion from the assumed linear graphene band structure, and is closely matched to the experimental results.  Electrical transport measurements for nearly aligned graphene on hBN devices also show sharp increases in resistance when the Fermi energy is tuned to the new superlattice Dirac points~\cite{Hunt:2013ef,Ponomarenko:2013hl,Dean:2013bv,Yang:2013ev}, providing further evidence of the reduced density of states there.

\subsubsection{Dirac point movement}

\begin{figure}
\begin{centering}
\includegraphics[width=3in]{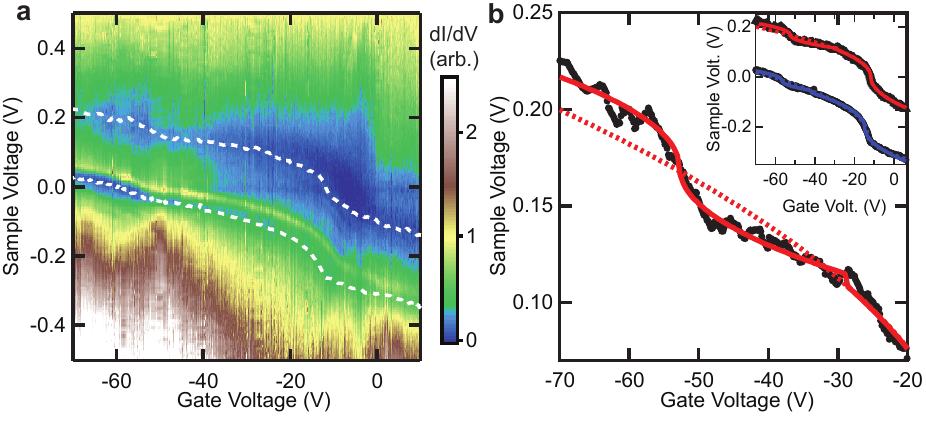}
\caption{(a) dI/dV as a function of sample and gate voltage for a 13.4 nm moir\'e pattern. White dotted lines mark the position of the original and superlattice Dirac points. (b) Energy of the Dirac point as a function of sample and gate voltage (black).  Solid red curve is a theoretical fit with the presence of the superlattice Dirac points.  Dotted red curve is the fit without the superlattice Dirac points.  The inset maps the position of the original and superlattice Dirac points and fits (red and blue, respectively) over a larger range of gate voltage. Reprinted with permission from Macmillian Publishers Ltd.: Yankowitz et al., Nature Physics \textbf{8}, 382 (2012), copyright 2012.}
\label{fig:GateSDP}
\end{centering}
\end{figure}

Figure~\ref{fig:GateSDP}(a) tracks the evolution of the LDOS as a function of gate voltage (which changes the chemical potential) for a 13.4 nm moir\'e pattern.  Two dips in the LDOS, highlighted by white dotted lines, move in parallel with changing gate voltage.  The upper dip is due to the original Dirac point, and the lower dip is due to the valence band superlattice Dirac point (the conduction band superlattice Dirac point is too weak to observe).  The parallel movement of these features provides evidence that the new dip in the LDOS of graphene on hBN is a feature of the band structure.  The gate dependence of the original Dirac point is given by the equation
\begin{equation}\label{eq:shift}
E_{\rm D} =\hbar v_{\rm F} \sqrt{2\alpha\pi(V_g - V_o)/g_v}
\end{equation}
where $V_g$ is the gate voltage, $V_0$ is the offset voltage, $\alpha$ is the coupling to the gate and $g_v$ is the valley degeneracy.  Near the original Dirac point ($\sim$-15 V in figure~\ref{fig:GateSDP}(a)), the valley degeneracy is $g_v$ = 2, and the resulting square root movement of the Dirac point gives a Fermi velocity of $v_{\rm F} = v_{\rm F}^0 = 0.94\pm0.02 \times 10^6$ m/s.  

When the Fermi energy is tuned near the superlattice Dirac point in the valence band ($\sim$-50V in figure~\ref{fig:GateSDP}(a)), the linearly vanishing density of states gives a second square root-dependent movement of the Dirac point, which does not occur for the case of bare graphene.  Here, the valley degeneracy is $g_v$ = 6 due to the three reciprocal superlattice vectors ${\bf G}_\alpha$ in each of the two Dirac cones.  The superlattice Dirac cones are predicted to be anisotropic with constant energy contours given by ellipses rather than circles.  Therefore, the Fermi velocity in Eq.~\ref{eq:shift} must be modified to $v_{\rm F} = \sqrt{v_{\rm F}^0v_{\rm F}^*}$ where $v_{\rm F}^0$ is the unmodified Fermi velocity parallel to ${\bf G}_{\alpha}$ and $v_{\rm F}^*$ is the reduced Fermi velocity perpendicular to ${\bf G}_{\alpha}$.  A fit of the movement of the Dirac point when the Fermi energy is tuned near the superlattice Dirac point gives a reduced Fermi velocity for the new electrons and holes of $v_{\rm F}^* = 0.5\pm0.1 \times 10^6$ m/s.  Figure~\ref{fig:GateSDP}(b) plots the movement of the Dirac point around that range of Fermi energies, and shows the second square root dependence of the Dirac point motion near the superlattice Dirac point.  The red dotted line plots the expected Dirac point movement for bare graphene.  Roughly speaking, the new superlattice Dirac cones are about twice as wide as the original Dirac cone.    

\subsubsection{Periodic charge variation}

\begin{figure}
\begin{centering}
\includegraphics[width=3in]{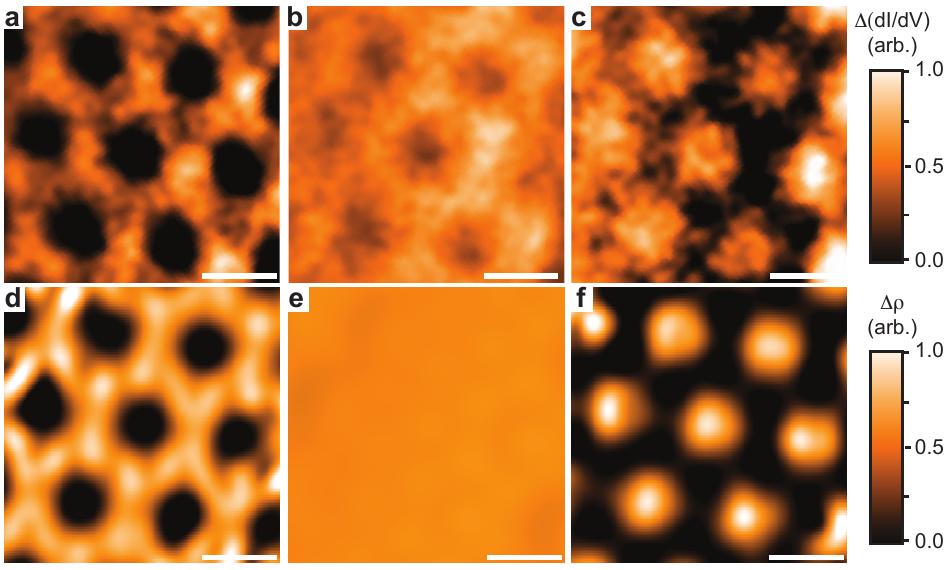}
\caption{(a) - (c) Experimental dI/dV maps for a 13.4 nm moir\'e pattern at sample voltages of -0.16 V, 0.17 V, and 0.44 V, respectively. (d) - (f) Theoretical dI/dV maps at energies of -0.3 eV, 0.03 eV, and 0.3 eV, respectively. Scale bar in all images is 10 nm. Reprinted with permission from Macmillian Publishers Ltd.: Yankowitz et al., Nature Physics \textbf{8}, 382 (2012), copyright 2012.}
\label{fig:SpatialSDP}
\end{centering}
\end{figure}

The presence of the periodic potential leads to a matching spatial variation in the LDOS~\cite{Yankowitz:2012gi,NeekAmal:2014ig}.  Theoretical calculations predict strong spatial modification in the LDOS at energies above (below) the conduction (valence) band superlattice Dirac point, but little modification at energies near the original Dirac point (figure~\ref{fig:SpatialSDP}(d) - (f)).  Furthermore, the relative strength of the hexagonal LDOS variation inverts between the valence and conduction bands.  In the valence band the perimeters of the hexagons exhibit an enhanced LDOS, whereas in the conduction band the centers of the hexagons exhibit an enhanced LDOS.  This hexagonal LDOS modulation spatially mirrors the periodic potential from the hBN substrate.  Figure~\ref{fig:SpatialSDP}(a) - (c) shows the corresponding experimental results, which exhibit behavior similar to the theoretical predictions.

\subsection{Other periodic potentials}\label{OtherPeriodic}

Similar superlattices can exist in graphene devices placed on other substrates~\cite{Wintterlin:2009bc,Li:2009eh,Rutter:2007epa,Gao:2010iz,Ndiaye:2006cr,Dedkov:2010jh,Kwon:2009kn,Sutter:2009ff,Roth:2011ja,Marchini:2007gt}.  The most noteworthy is graphene grown on metallic Ir(111).  The graphene layers grow with near perfect crystalline alignment with the underlying Ir, and due to the slight mismatch in lattice constants a 2.53 nm hexagonal moir\'e pattern is formed.  Angle resolved photoemission (ARPES) measurements have shown superlattice Dirac points similar to those in graphene on hBN~\cite{Pletikosic:2009fh}, although these samples lack the ability to tune the wavelength of the superlattice and thus the energy of these superlattice Dirac points.  Furthermore, due to the metallic nature of the Ir(111) substrate, gated transport measurements probing carrier dynamics in the graphene layer alone are not possible.  While measurements of graphene on other crystalline substrates have shown similar moir\'e patterns, they have yet to exhibit the existence of replica Dirac cones in the band structure, likely due to the coupling strength between the graphene and substrate~\cite{Pletikosic:2009fh}.

Twisted bilayer graphene exhibits similar moir\'e patterns in STM topography as well.  In this case, the lower graphene layer can be modeled as providing the periodic potential for the upper layer.  Twisted graphene bilayers are strongly coupled, and thus instead of opening superlattice Dirac points, the band structure consists of two neighboring Dirac cones at half-filling, connected by new van Hove singularities.  The energy of these van Hove singularities scales with the rotation between the layers.  Numerous STM studies have shown the existence of these van Hove singularities and explored their properties~\cite{Li:2009cp,Luican:2011hw,Brihuega:2012hh}.

\section{Unique electronic effects in graphene on hBN}\label{Unique}

In addition to the new electronic effects in graphene that are direct consequences of the hBN substrate, the cleaner charge environment and flat surface also allow the observation of intrinsic electronic properties of graphene which are normally obscured by the dirty and rough SiO$_2$ substrate.  For most rotation angles between the graphene and hBN lattices, the band structure renormalization from the hBN substrate is negligible at low energies and therefore does not interfere with the detection of graphene's intrinsic electronic properties.  

\subsection{LDOS oscillations}\label{Friedel}

For a uniform area of a given sample the electron wavefunctions can be represented as plane waves, therefore no spatial variation of the LDOS is expected.  However, defects in the sample can scatter electrons, creating standing waves which can be detected as oscillations in the LDOS.  The ``quantum corral''~\cite{Crommie:1993vk} experiment of 48 iron adatoms on a copper(111) surface is an excellent example, where scattering from a circle of adatoms on a metallic surface creates concentric rings of LDOS within the circle.

\subsubsection{Oscillation wavelength}

LDOS oscillations in monolayer graphene, as compared to those observed in noble metals, are about an order of magnitude longer in wavelength and decay more quickly.  These properties result from the peculiar band structure of graphene: the small diameter of Fermi circles explains the former feature, and the psuedospin quantum number the latter.  For graphene on SiO$_2$ samples, LDOS oscillations from impurities~\cite{Zhang:2009ce} or boundaries are obscured by the inhomogeneous charge environment making quantitative measurements of their decay difficult.  The cleaner charge environment provided by the hBN substrate permits analysis of such features.

Figure~\ref{LDOS} shows STS mapping near a step edge in a graphene on hBN sample~\cite{Xue:2012hd}. The step (dark horizontal band in the center of the images) is due to a layer number change in the underlying hBN.  The graphene layer on top conforms smoothly over the step.  LDOS oscillations parallel to the step emanate in both directions, though they show up more strongly in the lower half of these maps.  The wavelength of the oscillations increases as the tip bias is decreased from 138 meV in figure~\ref{LDOS}(a) to 38 meV in figure~\ref{LDOS}(f) (i.e. as the tip probes states progressively closer to the Dirac point).  The wavelength of these oscillations are on the order of 10 nm, whereas similar oscillations in noble metals are on the order of 1 nm~\cite{Crommie:1993co,Hasegawa:1993dq}. 

\begin{figure}
\begin{centering}
\includegraphics[width=3in]{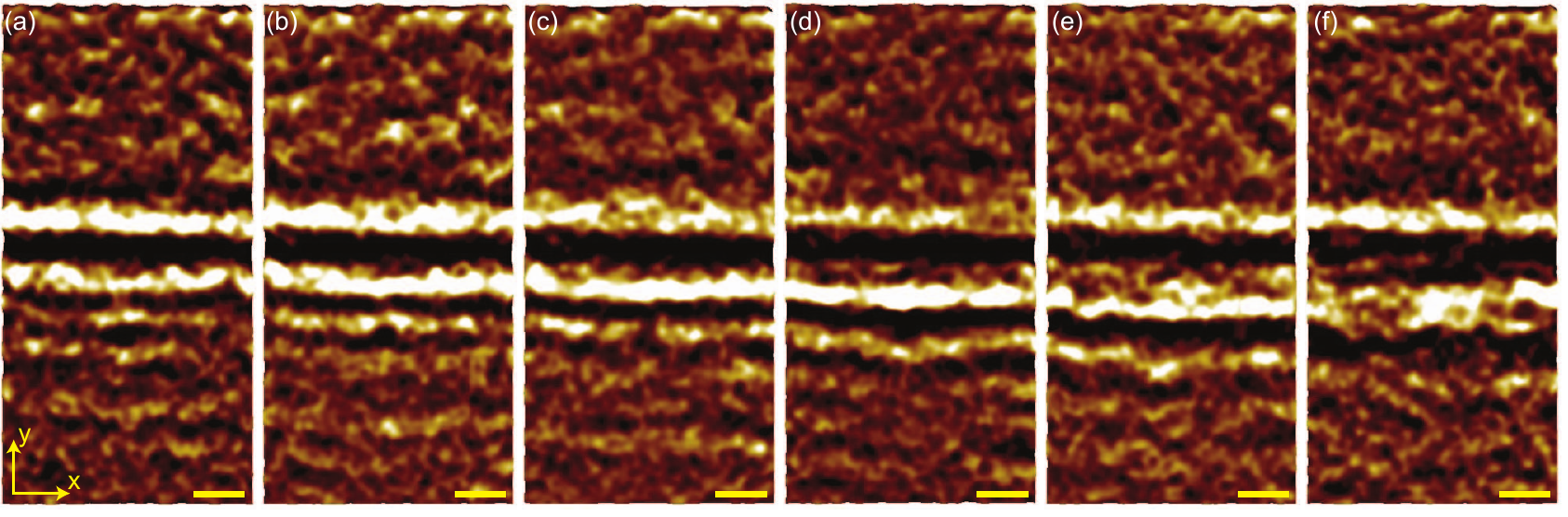}
\caption{(a) - (f) LDOS maps of graphene on hBN with an hBN step running horizontally.  Tip voltages are 138 mV, 118 mV, 98 mV, 78 mV, 58 mV, and 38 mV, respectively.  Scale bar is 10 nm in all images. Figure reproduced from Xue et al., Physical Review Letters \textbf{108}, 016801 (2012). Copyright 2012 by the American Physical Society.}
\label{LDOS}
\end{centering}
\end{figure}

\begin{figure}
\begin{centering}
\includegraphics[width=3in]{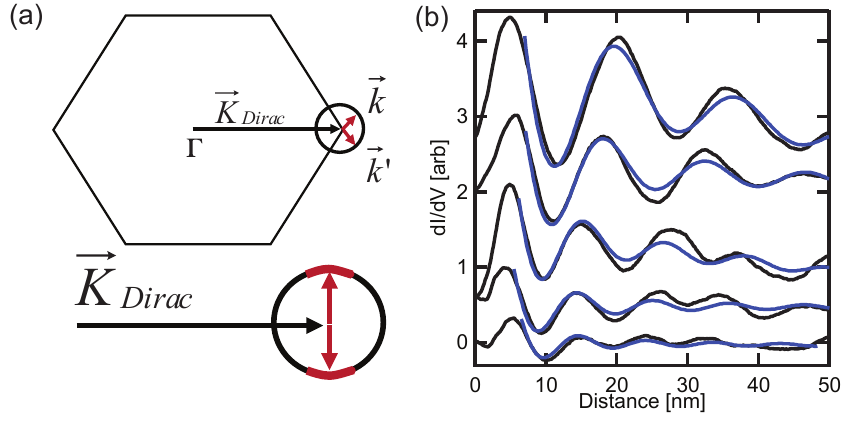}
\caption{(a) Schematic of scattering in graphene.  Momentum must be conserved along the direction of the edge $x$, and must change sign perpendicular to the edge.  (b) dI/dV at an hBN step edge taken at different tip voltages (black) and theoretical fits (blue). Figure adapted from Xue et al., Physical Review Letters \textbf{108}, 016801 (2012). Copyright 2012 by the American Physical Society.}
\label{LDOSfitting}
\end{centering}
\end{figure}

The LDOS oscillations result from electrons scattering at the step edge.  Electron waves propagating towards the step have some probability of transmission and reflection, depending on the angle of incidence.  The reflected portion of the electron wavefuction interferes with the incident wave and results in the observed LDOS standing waves.  The unusually long wavelength of these oscillations in graphene results from the peculiar nature of the graphene band structure.  As shown in figure~\ref{LDOSfitting}(a), when an electron with momentum ${\bf k}$ is scattered, due to conservation of energy and the momentum component parallel to the step, the resulting wave will have momentum ${\bf k'}$ within the constant energy circle centered in the same valley.  Since the step edge is not an atomic-scale defect, it cannot provide the large momentum transfer required for intervalley scattering.  The scattered electron wave with momentum ${\bf k'}$ interferes with the incident wave with momentum ${\bf k}$.  

The wavelength of the resulted standing wave can be calculated by considering ${\bf k}-{\bf k'}$, which is a small number since it is the result of an intravalley scattering process.  The small scattering wave vector corresponds to long wavelength LDOS oscillations.  Conversely, similar scattering on noble metal surfaces has much shorter wavelength LDOS oscillations.  Unlike graphene, which has small constant energy circles near its Fermi energy, noble metals usually have a large Fermi surface (large constant energy circles), so small momentum transfer scattering and hence long wavelength LDOS oscillations are not possible.  Figure~\ref{LDOSfitting}(a) also explains the observed behavior of the wavelength as a function of the energy shown in figure~\ref{LDOS}.  The diameter of the constant energy contour shrinks towards the Dirac point, thus ${\bf k}-{\bf k'}$ decreases and the wavelength of the LDOS oscillations increases.  Figure~\ref{LDOSfitting}(b) plots the line profiles of the oscillations (black curves) from the panels in figure~\ref{LDOS} (tip voltage increases from top to bottom).  The wavelength of the LDOS oscillations clearly increases as the energy approaches the Dirac point, as expected.

\subsubsection{Oscillation decay}
 
Additionally, the amplitude of the LDOS oscillations decreases as a function of distance from the step edge at all energies.  This feature can be explained by considering the relative contributions of incoming and reflected waves as a function of angle with respect to the step edge.  Pairs of incoming and reflected waves create oscillations with varying wavelengths as a function of angle.  Most of these contributions away from normal incidence tend to cancel each other far from the edge, resulting in a decreasing oscillation amplitude as a function of distance.  Similar decay behavior has been seen in noble metals~\cite{Crommie:1993co,Hasegawa:1993dq}.  However, the psuedospin quantum number forbids direct backscattering in graphene (an allowed process in noble metals)~\cite{Ando:1998fn}.  As a result, the LDOS oscillations decay faster than comparable oscillations in noble metals~\cite{Xue:2012hd,Zhou:2009fz}.  Theoretical calculations accounting for these factors (blue curves in figure~\ref{LDOSfitting}(b)) show good agreement with experimental measurements.

\subsection{Manipulation of atomic adsorbents}\label{Manipulation}

Coulomb potentials can be formed in graphene by depositing charged impurities on the graphene surface.  Work in this field has been done with cobalt, calcium and potassium atoms deposited {\it in situ} via electron beam evaporation~\cite{Ohta:2006bo,Brar:2011iq,Wang:2012id,Wang:2013ec}.  While it is possible to measure the electronic influence of these adatoms in graphene on SiO$_2$~\cite{Brar:2011iq}, the intrinsic electronic response of graphene to individual charged impurities can be obscured by the inhomogeneous charge environment.  The flatter surface of the hBN substrate allows for easier manipulation of adsorbed charged impurities and the cleaner charge environment allows interpretation of the intrinsic electronic response of graphene to these impurities.  Adatom monomers may be manipulated into dimers and trimers via diffusion or direct surface manipulation with the STM tip, and these larger charge impurities may be further manipulated on the surface with the STM tip as well. 

\subsubsection{Coulomb impurities}

\begin{figure}
\begin{centering}
\includegraphics[width=3in]{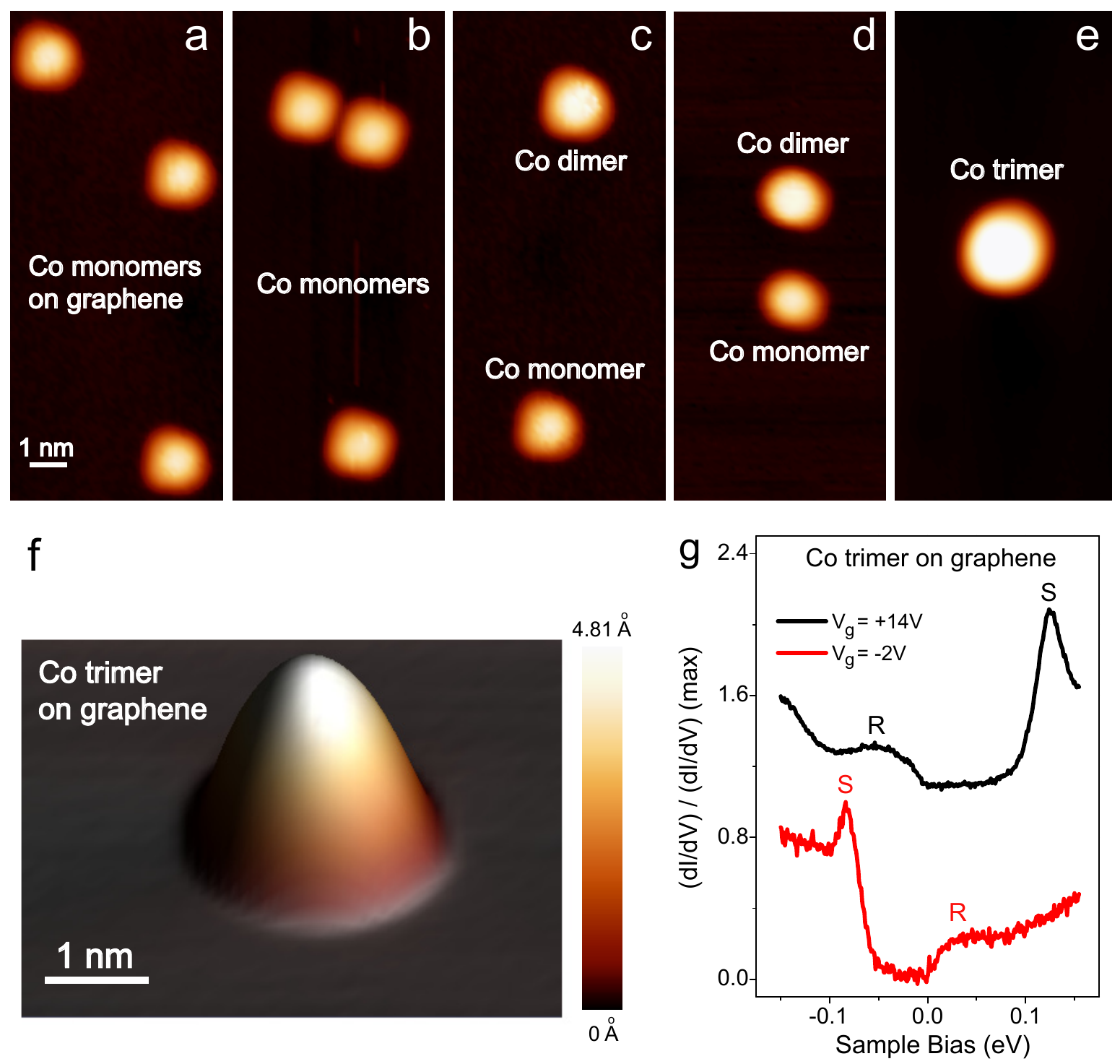}
\caption{(a) - (c) STM topography of graphene on hBN showing the manipulation of Co adatoms into a Co dimer using an STM tip.  (d),(e) Similar manipulation of a dimer and monomer into a Co trimer. (f) Zoomed in topography of the Co trimer. (g) dI/dV spectroscopy taken directly above the Co trimer at different back gate voltages. Reproduced with permission from Macmillian Publishers Ltd.: Wang et al., Nature Physics \textbf{8}, 653 (2012), copyright 2012.}
\label{fig:TrimerManipulation}
\end{centering}
\end{figure}

Wang et al. explored the electronic effects of a single Coulomb impurity on graphene~\cite{Wang:2012id}.  Coulomb impurities are formed via electron beam evaporation of Co atoms onto the surface of graphene.  These surface adatoms are manipulated into Co trimers on the atomically flat and clean graphene on hBN surface using an STM tip.  The trimer configuration is chosen as it is the fewest adatom system which exhibits charge toggling via a back gate electrode and charge state stability in the presence of an STM tip.  Figure~\ref{fig:TrimerManipulation}(a) - (e) shows the process of manipulating Co monomers on graphene on hBN into a Co trimer, while figure~\ref{fig:TrimerManipulation}(f) shows a zoomed-in topography image of the Co trimer.  The dI/dV spectroscopy (figure~\ref{fig:TrimerManipulation}(g)) taken with the tip directly above the Co trimer exhibits the characteristic {\bf R} and {\bf S} states previously found in Co monomers on graphene on SiO$_2$~\cite{Brar:2011iq}.  The {\bf R} state arises from local cobalt-graphene hybridization, and the {\bf S} state is due to tip-induced ionization of the trimer.

\begin{figure}
\begin{centering}
\includegraphics[width=3in]{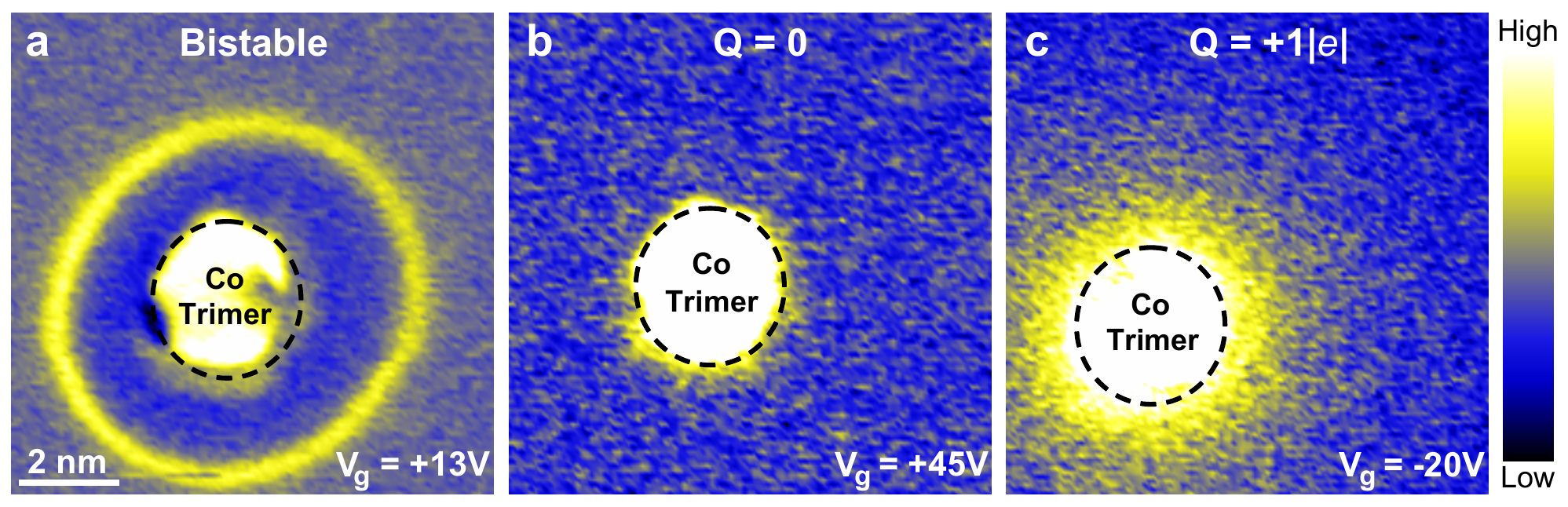}
\caption{dI/dV map over a Co trimer in the (a) bistable, (b) Q = 0, and (c) Q = +1$|e|$ charge states. Reproduced with permission from Macmillian Publishers Ltd.: Wang et al., Nature Physics \textbf{8}, 653 (2012), copyright 2012.}
\label{fig:TrimerMap}
\end{centering}
\end{figure}

Figure~\ref{fig:TrimerMap} shows maps of the dI/dV spectroscopy response of graphene to a Co trimer in three different charge states at a sample voltage of +0.3 eV, achieved by varying the back gate voltage.  The clean charge environment from the hBN substrate ensures the spectroscopy response represents the intrinsic electronic behavior of graphene.  The combination of the tip and back gate is capable of charging the Co trimer by one electron.  For large positive gate voltages (figure~\ref{fig:TrimerMap}(b)) the Co trimer is always uncharged, and for large negative gate voltages (figure~\ref{fig:TrimerMap}(c)) the Co trimer is always charged with one electron for all positions of the tip.  At intermediate gate voltages (figure~\ref{fig:TrimerMap}(a)) the Co trimer is in a bistable configuration, where the charge of the trimer can be changed by one electron depending on whether the tip is inside or outside the ring feature surrounding the trimer.      

\subsubsection{Atomic collapse} 

Among its many incredible properties, graphene can serve as a low-energy testbed for mimicking high-energy phenomena, such as Klein tunneling~\cite{Katsnelson:2006kd,Stander:2009ce,Young:2009il} and atomic collapse~\cite{Wang:2013ec}.  The latter is a prediction that highly charged atomic nuclei are unstable due to relativistic quantum effects.  Specifically, if the nuclear charge Z exceeds a critical value Z$_c$, the electron wavefunction falls towards the nucleus and the positron component escapes to infinity~\cite{Darwin:1913gv,Greiner:1985ij,Boyer:2004hu,Zeldovich:2007jk}.  In contrast, subcritical nuclei have stable atomic bound states.  For real atoms, Z$_c$ $\sim$ 170 can only be achieved through colliding heavy atoms, and interpretation of the results is challenging.  Results from these efforts remain ambiguous~\cite{Schweppe:1983fm,Cowan:1985iu}.  However, in graphene Z$_c$ is instead of the order one due to the relativistic nature of the charge carriers, as well as the large effective fine-structure constant~\cite{Pereira:2007di,Shytov:2007ck,Shytov:2007gu}.  In this case, holes play the role of positrons.  Near ``artificial nuclei'' on graphene of charge greater than Z$_c$, a spatially extended electronic resonance corresponding to the electron-like part of the wave function is expected just below the Dirac point.  Ref.~\cite{Wang:2013ec} explores this prediction by assembling artificial nuclei of different charge via spatial manipulation of adsorbed Ca dimers.

\begin{figure}
\begin{centering}
\includegraphics[width=3in]{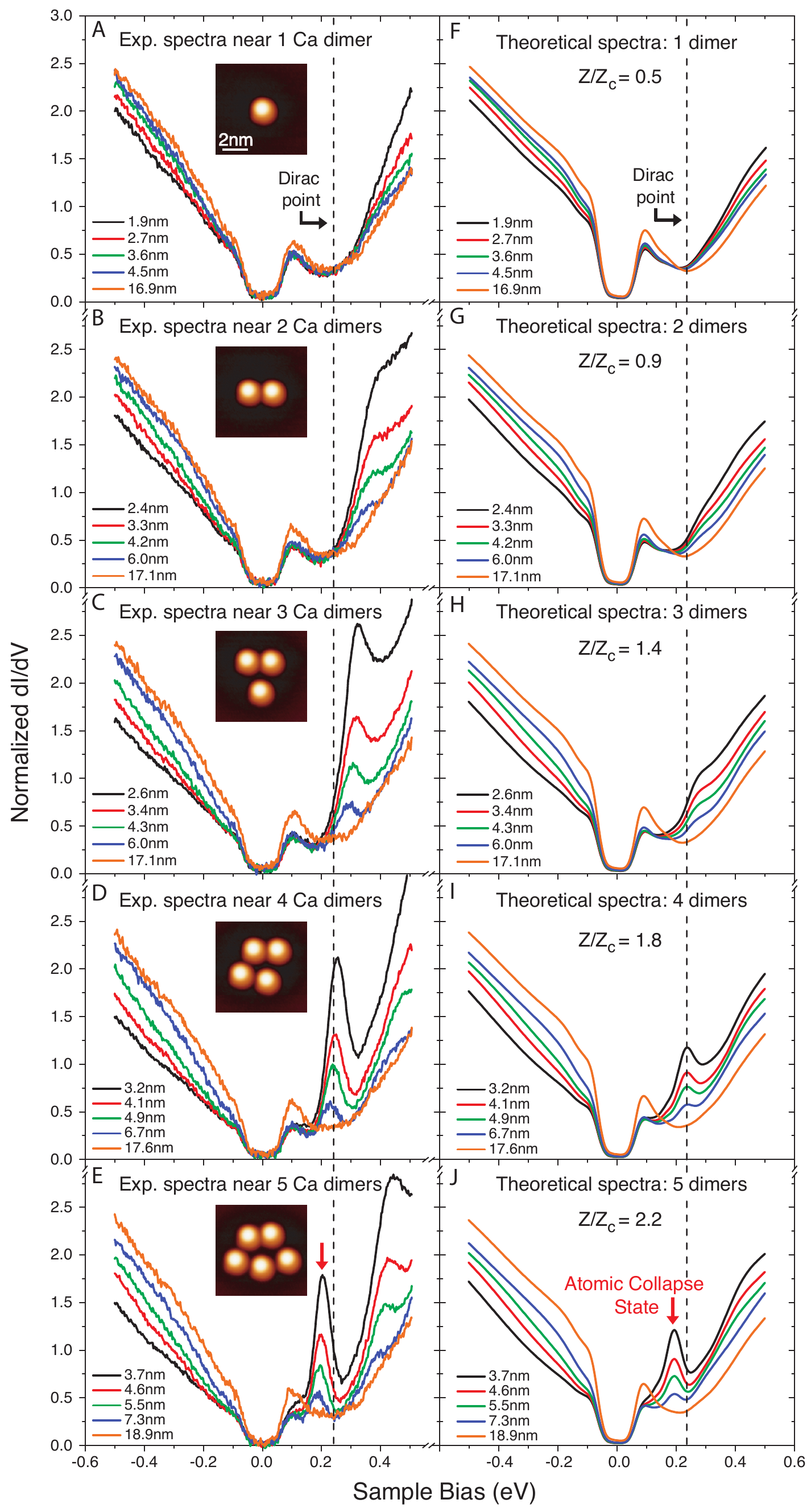}
\caption{(a) - (e) dI/dV spectroscopy taken at varying distances from the center of Ca clusters. Insets show topography of the Ca clusters and indicate the number of adatoms in the cluster. (f) - (j) Theoretical dI/dV predictions for comparable measurements.  The black dotted line marks the Dirac point energy in all panels.  Red arrows indicate the atomic collapse state. From Wang et al., Science \textbf{340}, 6133 (2013). Reprinted with permission from AAAS.}
\label{fig:AtomicCollapse}
\end{centering}
\end{figure}

The insets of figures~\ref{fig:AtomicCollapse}(a) - (e) show STM topography of clusters of Ca dimers ranging from one to five dimers arranged via an STM tip on the surface of graphene on hBN.  The main panels show dI/dV spectroscopy taken at a varying range of distances from the cluster center.  As the number of dimers in the cluster is increased a bound-state resonance develops near the Dirac point, and drops below the Dirac point energy for a cluster of five dimers.  This resonance represents the atomic collapse state.  Despite the spatial separation and asymmetric placement of these dimers, spatial maps of the quasi-bound state resonance intensity in the region surrounding the dimer cluster is highly symmetric and extends greater than 10 nm from the cluster center.  This provides evidence that the dimer cluster behaves as a single artificial nucleus, and that the graphene charge carriers are spatially separated from the dimers.

Figures~\ref{fig:AtomicCollapse}(f) - (j) plot simulated dI/dV spectroscopy next to the corresponding experimental data.  The simulations are calculated assuming a two-dimensional continuum Dirac model for graphene in the presence of a Coulomb potential.  The only essential fitting parameter is Z/Z$_c$.  Z is the effective screened charge seen by the graphene charge carriers.  Z$_c$ in graphene is $Z_{\rm c} = \hbar v_{\rm F}/(2 e^2)$ $\sim$ 0.25.  The best fit values for Z/Z$_c$ are given in figures~\ref{fig:AtomicCollapse}(f) - (j).  

\subsection{Magnetic field effects}\label{Magnetic}

Local measurements of graphene on hBN in a magnetic field show very well developed Landau levels owing to the reduced charge environment due to the hBN substrate~\cite{Chae:2012jk,Zou:2013dx,LuicanMayer:2014jt,Kou:2013ud}.  Furthermore, new many-body effects, which were previously smeared in graphene on SiO$_2$, are accessible in the cleaner graphene on hBN samples.  

\begin{figure}
\begin{centering}
\includegraphics[width=3in]{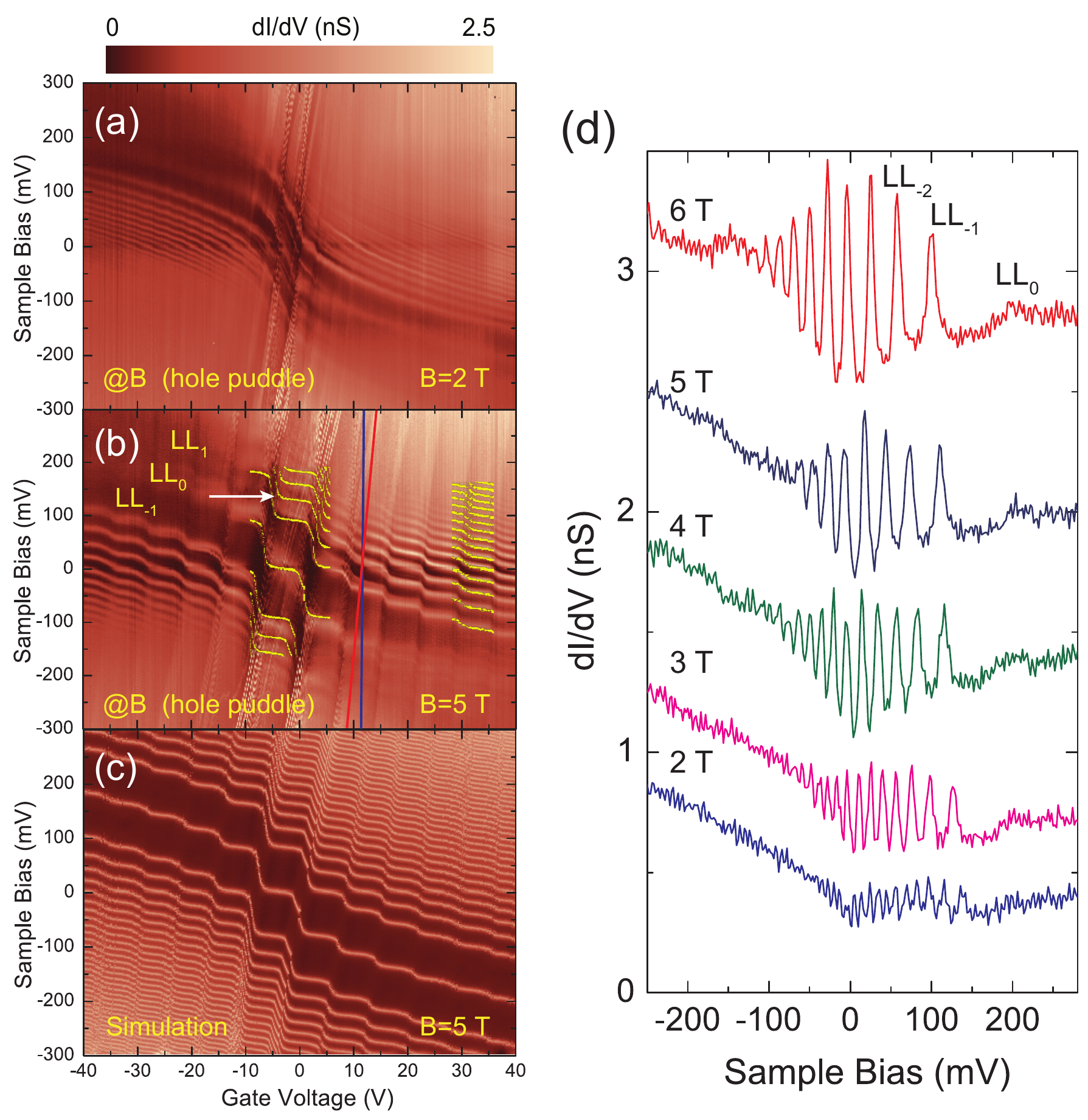}
\caption{(a) dI/dV spectroscopy as a function of sample and gate voltage at B = 2 T. (b) Same at B = 5 T.  The blue line denotes constant gate voltage and the red line denotes constant charge density.  The yellow lines are theoretical predictions from the simulation in (c). (c) Simulation of the data shown in (b). (d) dI/dV spectroscopy of graphene on hBN in varying magnetic fields showing well developed Landau levels. Figure adapted from Chae et al., Physical Review Letters \textbf{109}, 116802 (2012). Copyright 2012 by the American Physical Society.}
\label{fig:LLGate}
\end{centering}
\end{figure}

Ref.~\cite{Chae:2012jk} provides a study of graphene on hBN in magnetic fields from 0 to 8 T.  The LLs are extremely sharp and are well developed at fields as small as 2 T, as shown in the dI/dV spectroscopy in figure~\ref{fig:LLGate}(d).  The energy of the LLs as a function of magnetic field behaves as expected for bare graphene, where the linear dispersion gives E$_N$ = E$_D$ + $sgn(N) \sqrt{2 e \hbar  v_{\rm F}^2 N B}$, where E$_D$ is the energy of the Dirac point.

Figures~\ref{fig:LLGate}(a) and (b) track the dI/dV spectroscopy as a function of gate voltage and magnetic field.  The LLs become sharper and more step-like with increasing magnetic field.  The staircase pattern results from partially-filled LLs being pinned at the Fermi energy.  A quick transition to the next LL is made when the prior LL is filled.  The overall behavior of the gate dependent spectroscopy is consistent with the theoretical simulation shown in figure~\ref{fig:LLGate}(c).  Furthermore, at large gate voltages, and thus large charge carrier density, the simulation provides a good quantitative fit to the experimental results as evidenced by the yellow theory curves overlayed above the experimental data.  However, at low carrier densities the simulated peak positions are poorly matched to the experimental data.

\begin{figure}
\begin{centering}
\includegraphics[width=3in]{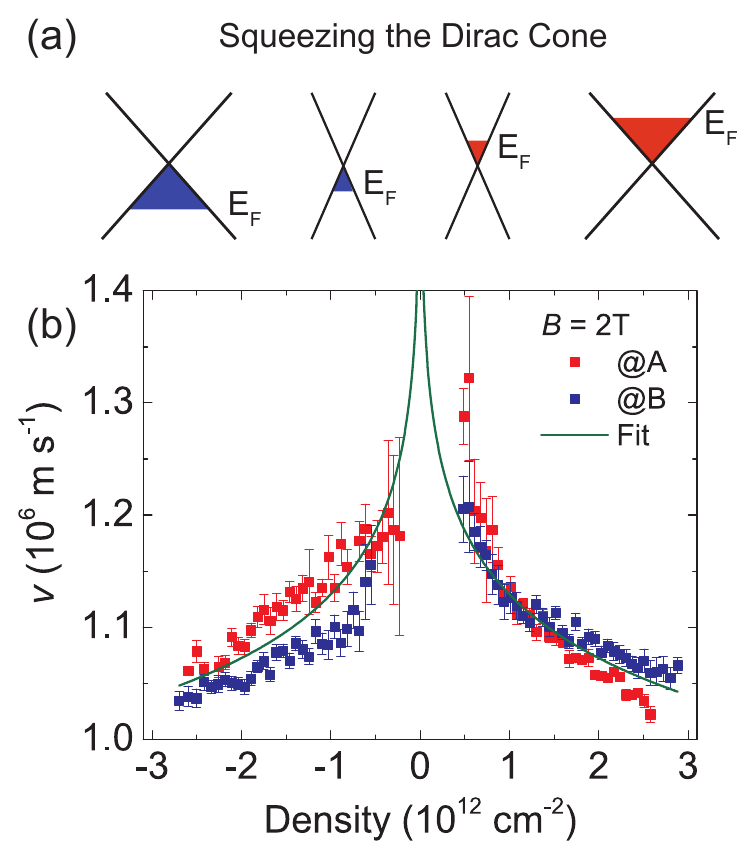}
\caption{(a) Schematic of the Dirac cone as a function of charge density.  The cone squeezes (the velocity rises) as the density decreases. (b) Experimentally found renormalized velocity as a function of density at B = 2 T.  Red (blue) symbols are taken in an electron (hole) puddle.  The green line is a theoretical fit of the data. Figure reprinted from Chae et al., Physical Review Letters \textbf{109}, 116802 (2012). Copyright 2012 by the American Physical Society.}
\label{fig:FermiVRenorm}
\end{centering}
\end{figure}

The low density deviations provide evidence for the breakdown of the single particle behavior assumed in the theoretical simulation.  This result shows the importance of accounting for many-body interactions at low densities.  Plotting E as a function of $\sqrt{NB}$ for different carrier densities reveals that the LL dispersion remains linear even at very low densities, however the slope changes as a function of carrier density.  This implies that while the graphene energy dispersion always remains linear, the Dirac cone is squeezed due to interaction effects at low densities (figure~\ref{fig:FermiVRenorm}(a)).  Figure~\ref{fig:FermiVRenorm}(b) plots the Fermi velocity fit over a range of carrier densities for both an electron and hole doped region.  Both indicate a renormalized Fermi velocity at low carrier densities, with measured low density values over 30\% larger than the bare graphene velocity.  Transport measurements of suspended graphene devices have revealed similar physics, although without the ability to tune excitation energy and carrier density separately~\cite{Elias:2011ev}.

\section{Electrical transport measurements}\label{Transport}

In prior sections we discussed, from the perspective of STM and STS measurements, the benefits of an hBN substrate for graphene devices and the new physics which becomes accessible as a result.  In the following section we highlight a few of the recent transport experiments that require either the ultra low charge variation provided by the hBN substrate or the superlattice Dirac points created by the periodic potential.  Important insights into the novel physics observed in these transport experiments have come directly from results of STM measurements of graphene on hBN. 

\subsection{Integer quantum hall effect}\label{IQHE}

\begin{figure}
\begin{centering}
\includegraphics[width=3in]{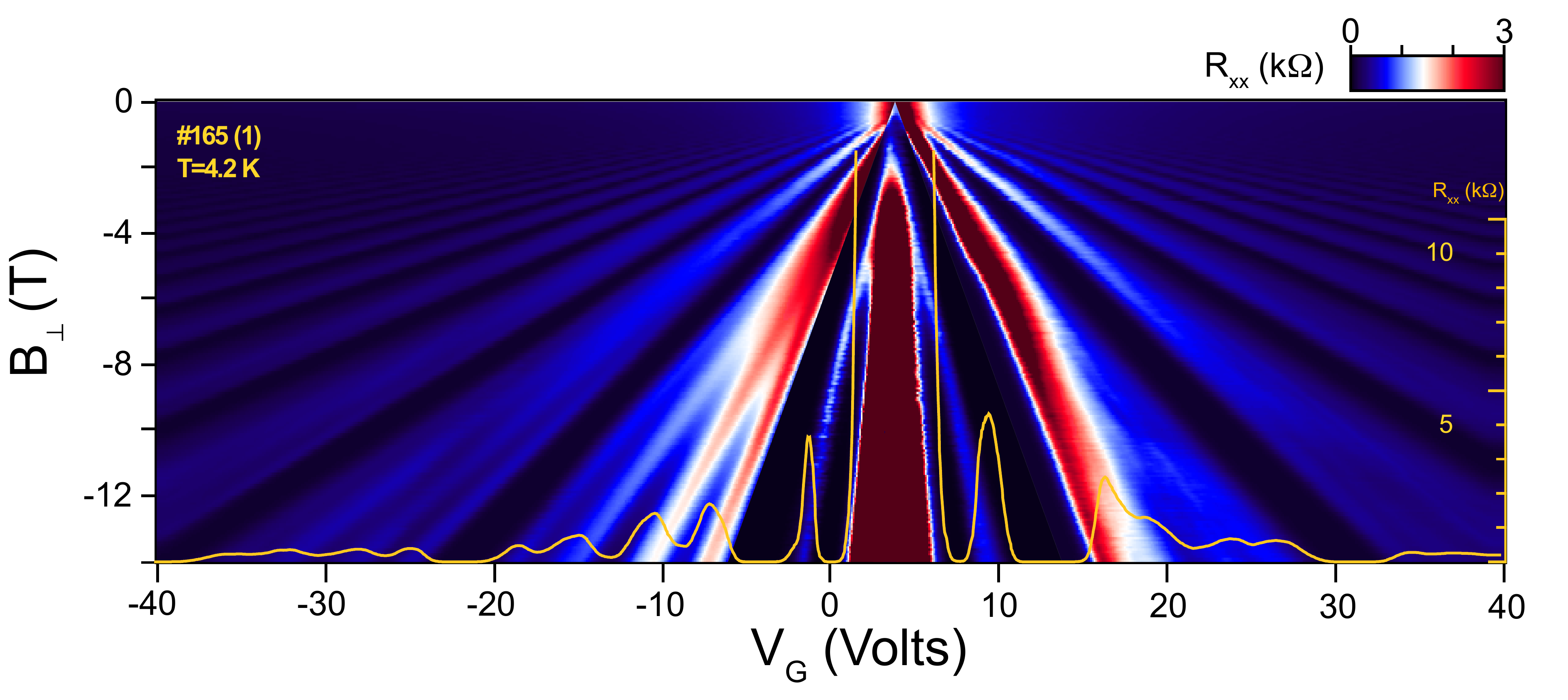}
\caption{Landau fan of graphene on hBN.  The superimposed curve is a cut at B = 14 T, exhibiting four-fold symmetry breaking.  Adapted with permission from Macmillian Publishers Ltd.: Young et al., Nature Physics \textbf{8}, 550 (2012), copyright 2012.}
\label{fig:IQHE}
\end{centering}
\end{figure}

Due to the spin and valley degeneracy of graphene, each Landau level is four fold degenerate (in bilayer graphene, the lowest energy Landau level has an extra four fold degeneracy).  This provides an ideal test bed for exploring the rich physics of multicomponent quantum Hall effects~\cite{DasSarma:1997hr}.  In the early transport experiments where SiO$_2$ was the common choice of substrate, quantum Hall measurements showed Hall conductance plateaus at $\sigma_{xy}=\pm(N+1/2)4e^2/h$ for monolayer graphene and $\sigma_{xy}=\pm 4Ne^2/h$ ($N\neq0$) for bilayer graphene~\cite{Novoselov:2005es,Zhang:2005gp,Novoselov:2006hu}.  Broken symmetry effects such as Zeeman splitting, long range Coulomb interaction, and lattice scale electron-electron interactions have been shown to partially lift the degeneracy in suspended graphene~\cite{Du:2009ce,Bolotin:2009ko} and graphene on SiO$_2$~\cite{Zhang:2006hn} under high magnetic field. However, due to the fragile device structure of suspended samples and the dirty charge environment of the SiO$_2$ substrate, studying broken symmetry states in these samples is challenging.  With hBN as the substrate, these states can be readily observed and the full four-fold degeneracy can be easily lifted at lower magnetic fields, hence all the integer filling factors can be observed~\cite{Dean:2010jy}.  Figure~\ref{fig:IQHE} shows the LL development with increasing magnetic field for graphene on hBN.  The darkest regions of the LL fan correspond to the N = 0, 1, 2, etc. LLs, while the four-fold broken symmetry states begin to emerge at fields above B = 7 T. 

To understand the origin of the broken symmetry states, Young et al. performed tilted magnetic field measurements of graphene on hBN devices~\cite{Young:2012bn}.  In this type of measurement, the sample is tilted at different angles relative to the magnetic field.  The overall field strength is changed to keep the perpendicular component of the field constant.  Since graphene is only one atom thick, the in-plane field couples to the system only through the electron spin.  By measuring how the energy gaps of different quantum Hall states evolve as a function of in-plane field (or total field, since the perpendicular field is fixed), one can gain important insights into the spin structure of the broken symmetry states. 

With this technique, Young et al. found that the symmetry broken order of the zeroth Landau level is different from higher Landau levels.  At filling factor $\nu=0$, the zeroth Landau level is half filled, and an insulating state is present.  The resistance of this state is found to decrease with increasing total magnetic field, which signifies that the ground state at $\nu=0$ is not spin polarized, instead, it has a broken valley symmetry origin.  Therefore, the ground state is valley polarized.  On the other hand, energy gaps at half fillings of higher Landau levels, such as $\nu=4, 8, 12$, etc. increase with total magnetic field, indicating that the ground states are spin polarized. To further confirm these broken symmetry assignments, Young et al. examined energy gaps at quarter fillings, such as $\nu=-1$ from the zeroth Landau level and $\nu=-3$ and $\nu=-5$ from the first Landau level in the hole side. They found that as expected, $\nu=-1$ state is spin polarized as the gap size increases with total magnetic field, while $\nu=-3$ and $\nu=-5$ gaps in general show minimal dependence on the total magnetic field.  Behavior deviating from this was also observed in some samples studied, indicating more work is necessary to gain a full understanding of the underlying physics.

\subsection{Fractional quantum hall effect}\label{FQHE}

\begin{figure}
\begin{centering}
\includegraphics[width=3in]{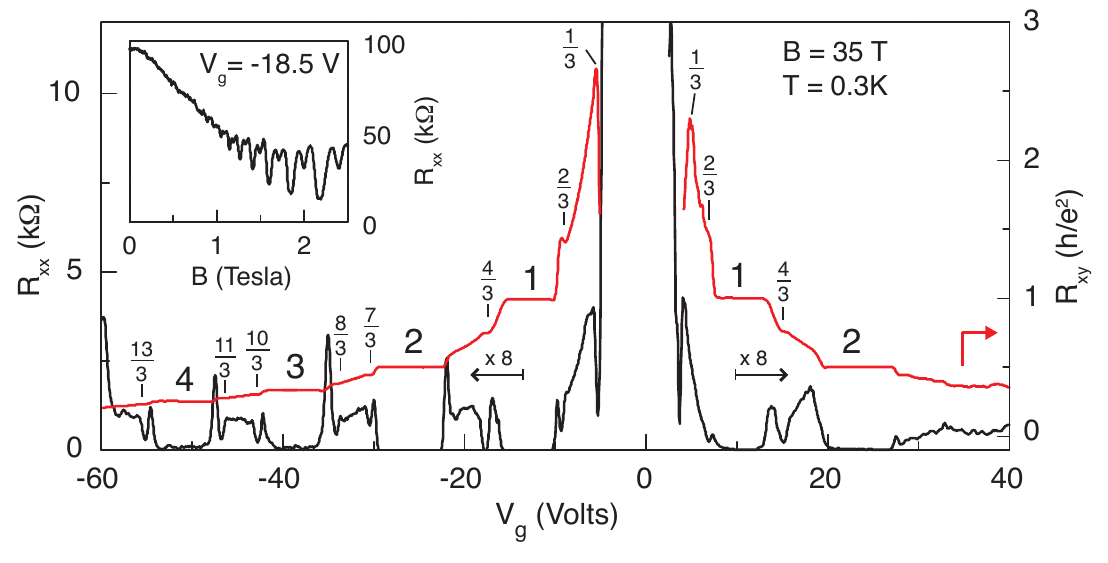}
\caption{Magnetoresistance (black) and Hall resistance (red) as a function of gate voltage at B = 35 T.  Adapted with permission from Macmillian Publishers Ltd.: Dean et al., Nature Physics \textbf{7}, 693 (2011), copyright 2011.}
\label{fig:FQHE}
\end{centering}
\end{figure}

The fractional quantum Hall effect has been observed by Dean et al.~\cite{Dean:2011ks} with high quality graphene on hBN samples.  Figure~\ref{fig:FQHE} illustrates magnetoresistance and Hall resistance results in a magnetic field of B = 35 T.  Previously, this many-body interaction effect could only be observed in high quality suspended devices~\cite{Du:2009ce,Bolotin:2009ko}. Compared with the suspended structure, hBN supported devices are more robust and easier to electrically contact and measure.  Therefore, it is possible to make multi-terminal devices on hBN substrates.  Another advantage of the supported structure is that higher carrier density can be achieved by applying larger gate voltages, which is problematic for suspended devices because they tend to break.

In their devices, Dean et al. found quantum Hall states at fractional filling of 1/3 and other equivalent states, such as 2/3 and 4/3 in the $\nu=0$ and $\nu=1$ Landau levels, respectively.  The presence of these states can be understood within the picture of composite fermions~\cite{Laughlin:1983hk,Stormer:1999iz}.  The absence of the 5/3 state in these devices is intriguing, since it should be the closest analogue to the 1/3 state in conventional semiconductors.  The absence of the 5/3 state is presumably due to some remaining symmetry from the original four fold degeneracy.  In contrast, all multiples of the 1/3 fractions are observed in the second LL.  This may signify the importance of density dependent electron-electron interactions. 

\subsection{Double layer devices}\label{Double}

\begin{figure}
\begin{centering}
\includegraphics[width=3in]{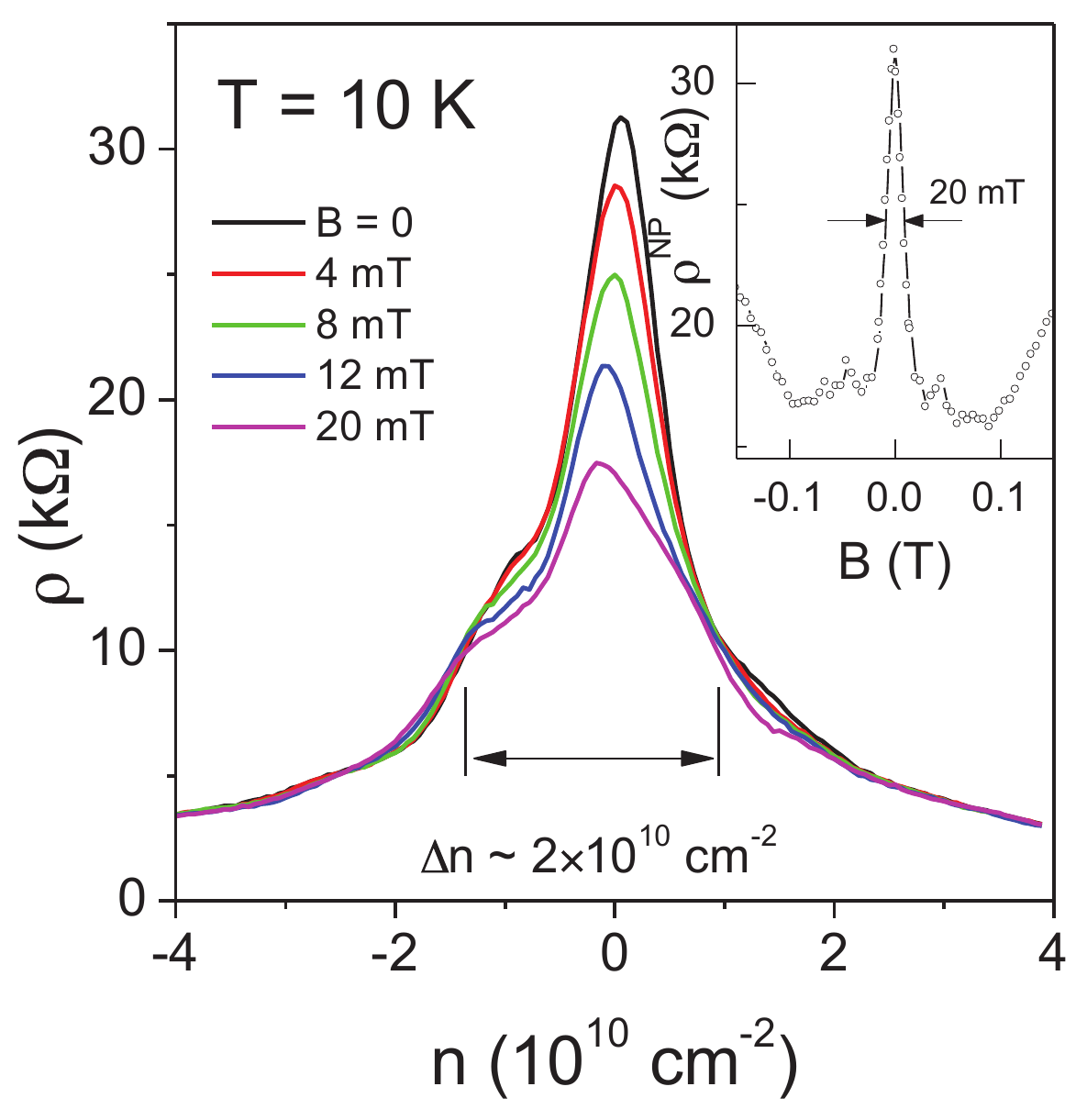}
\caption{Insulating state resistance as a function of magnetic field at T = 10 K.  The inset plots the resistance of the charge neutrality point as a function of magnetic field. Reproduced with permission from Macmillian Publishers Ltd.: Ponomarenko et al., Nature Physics \textbf{7}, 958 (2011), copyright 2011.}
\label{fig:Anderson}
\end{centering}
\end{figure}

Graphene on hBN heterostructures can be repeated vertically to form multilayered devices.  The simplest version of this type of structure is two layers of graphene separated by hBN.  Ponomarenko et al. showed that with high carrier density (n $>$ 10$^{11}$ cm$^{-2}$) induced in the bottom graphene layer, the top graphene layer shows a diverging resistance at the charge neutrality point~\cite{Ponomarenko:2011cj}.  Interestingly, this diverging resistance can be strongly suppressed by a small perpendicular magnetic field B $<$ 0.1 T (figure~\ref{fig:Anderson}), a signature of anti-localization.  The authors reasoned that the bottom layer graphene with high charge density screens the electrostatic potential from the SiO$_2$ substrate.  Therefore, the electron-hole puddles in the top graphene layer are further reduced compared with simple graphene on hBN devices. With this further improved charge homogeneity, a very low charge carrier density of $\sim$10$^{10}$ cm$^{-2}$ can be reached and the graphene resistance reaches the threshold value for localization, i.e. h/4e$^2$, where the factor of four accounts for the doubly degenerate spin and valley quantum numbers.  However, this interpretation of the result is still under debate~\cite{Amet:2013gw,Kechedzhi:2012ij} and a complete understanding of this phenomena is still lacking.

A similar device structure has also been measured by Hunt et al., where the graphene on hBN heterostructure sits atop a thick piece of graphite, which is contacted separately to act as a back gate~\cite{Hunt:2013ef}.  This device geometry provides an even flatter and cleaner environment for the top layer of graphene than devices using a single layer of graphene as the substrate for the graphene on hBN heterostructure, as the thick graphite gate is able to better screen charge inhomogeneities in the SiO$_2$.  These devices have exhibited full spin and sublattice symmetry breaking in the graphene, as evidenced by the presence of the $\nu=5/3$ fractional quantum Hall state.  They also exhibit a band gap at the charge neutrality point, whose size increases with moir\'e wavelength.  Recent theoretical studies have worked to explain these phenomena~\cite{Song:2013ji,Bokdam:2014vm}.  STM topography was used in this experiment to confirm the very long moir\'e wavelengths determined by transport measurements, as well as to measure the shorter moir\'e wavelengths which were inaccessible for the range of carrier densities probed. 

\subsection{Hofstadter butterfly}\label{Butterfly}

\begin{figure}
\begin{centering}
\includegraphics[width=3in]{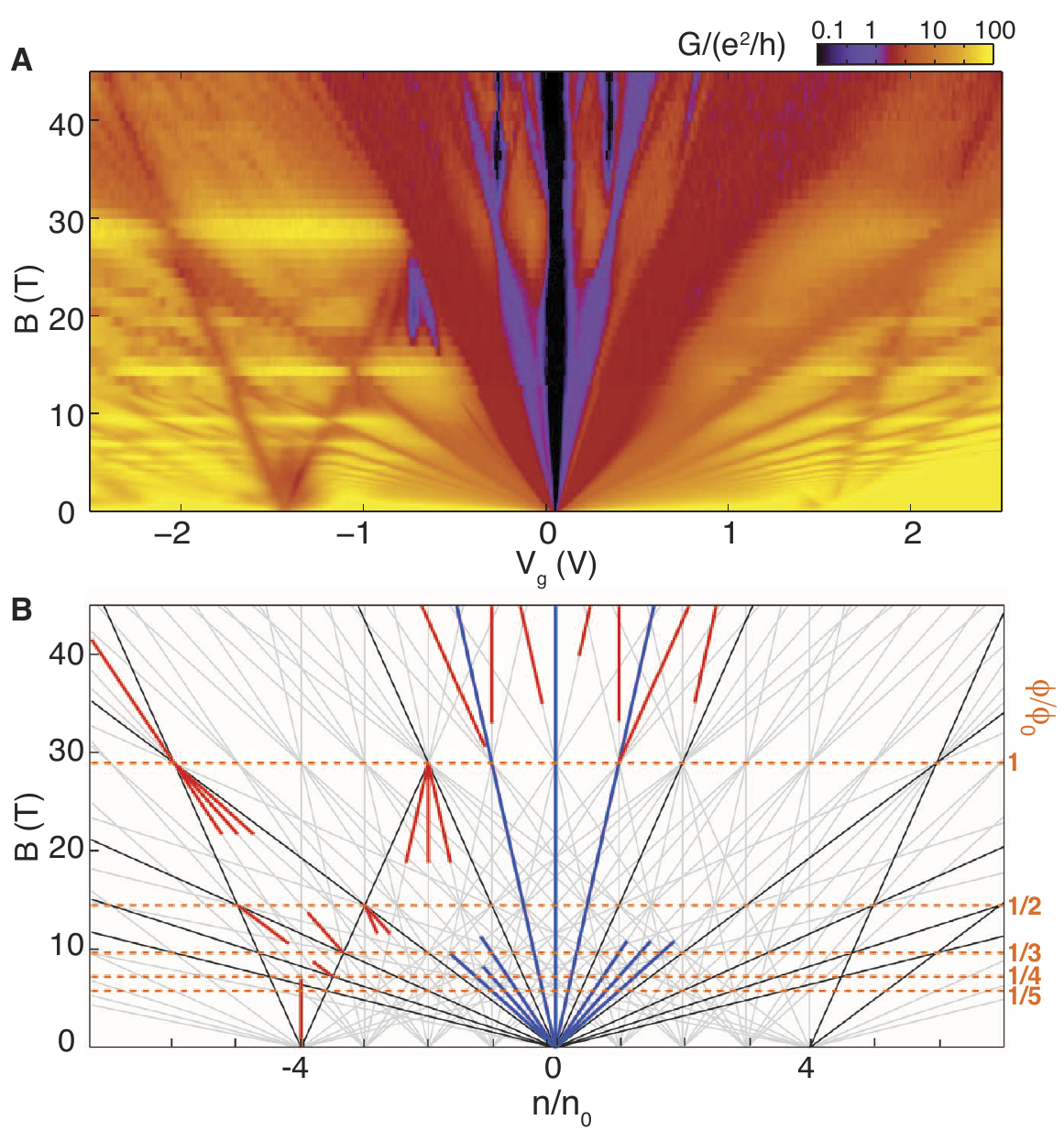}
\caption{(a) Magnetoconductance of graphene on hBN with near perfect rotational alignment.  Landau level fans originate from the original Dirac point (around V$_g$ = 0 V) and from the superlattice Dirac points (around V$_g$ = $\pm$1.5 V). (b) Theoretical calculation similar to (a).  Black lines represent gaps requiring no broken symmetry. Blue and red are broken symmetry states of the original and superlattice Landau fans, respectively. Gaps intersect for $\phi/\phi_0=1/q$, where $q$ is an integer. From Hunt et al., Science \textbf{340}, 6139 (2013). Reprinted with permission from AAAS.}
\label{fig:Butterfly}
\end{centering}
\end{figure}

Probing the behavior of superlattice Dirac points in a magnetic field proves challenging for most STM experiments, where only modest magnetic fields may be applied.  Transport experiments can be used in conjunction with STM results to probe the interplay between periodic potentials and magnetic fields.  Long wavelength superlattice potentials found in nearly aligned graphene on hBN samples provide a unique platform for studying charge carriers where the wavelength of the periodic potential is on the order of the magnetic length.  As outlined in \Sref{LL in magnetic field}, the unusual energy level dispersion of LLs in graphene follows E$_N$ = $sgn(N) \sqrt{2 e \hbar v^2 N B}$, with E$_N$ referenced to the Dirac point energy.  Superlattice Dirac points exhibit their own electron and hole carriers, whose energies also break into Landau levels with a similar dispersion in a magnetic field.  Transport experiments have shown Landau fans originating from both the original and superlattice Dirac points~\cite{Hunt:2013ef,Ponomarenko:2013hl,Dean:2013bv,Yang:2013ev}.  At low magnetic field, Landau fans originating from the superlattice Dirac points are only accessible through back gate tuning for very long wavelength moir\'e patterns, where the superlattice Dirac points are closest to the original Dirac point.  As the magnetic field is increased, Landau levels from the original and superlattice bands intersect for $\phi/\phi_0=1/q$, where $\phi$ is the flux through one superlattice unit cell, $\phi_0$ is the flux quantum and $q$ is an integer.  The resulting Landau levels create a recursive structure known as the ``Hofstadter butterfly'' spectrum~\cite{Hunt:2013ef,Ponomarenko:2013hl,Dean:2013bv,Hofstadter:1976js}.  Long wavelength periodic potentials formed by graphene on hBN allow the observation of this Hofstadter butterfly spectrum at magnetic fields accessible in the lab, which must be able to fill more than one flux quantum quantum per superlattice unit cell.  Figure~\ref{fig:Butterfly}(a) shows one such measurement for a nearly perfectly aligned graphene on hBN device.   Figure~\ref{fig:Butterfly}(b) provides a theoretical calculation for the system, indicting which broken symmetry states of the original and superlattice Dirac points are found experimentally.

\section{Conclusion}\label{Conclusion}

We have reviewed the properties of graphene on hBN devices from the perspective of scanning tunneling microscopy.  Graphene on hBN provides, to date, the most convenient structure for achieving clean graphene devices.  Graphene on hBN devices show charge impurity densities approaching suspended graphene devices, but are much easier to fabricate and contact electrically, and are much more physically robust.  Furthermore, the hBN substrate acts as a periodic potential for graphene, opening new Dirac points in the electronic band structure which may be exploited to create novel devices.  Scanning tunneling microscopy is an ideal tool for understanding the improvement of the hBN substrate over the standard SiO$_2$ substrate as it provides a direct probe of the charge landscape in the graphene.  Furthermore, STM provides direct experimental access to local electronic properties of graphene on hBN which are much more difficult or impossible to observe in graphene on SiO$_2$ devices.  These results can be used to aid the understanding of novel global transport experiments which exploit the benefits of the hBN substrate.  

Graphene on hBN heterostructures represent an important family of devices which will be used for the discovery of new physical phenomena and the development next-generation electronics featuring designer electrical properties.  With greatly improved transfer techniques~\cite{Wang:2013ch,Zomer:2014wu} which never require polymers or chemicals to touch the surfaces of the 2D materials, device cleanliness is nearing its ultimate limit.  Consequently, the electronic properties of such heterostructures may be studied in the ballistic regime, in the absence of virtually all disorder.  Adding other van der Waals materials to these heterostructures~\cite{Geim:2013dq} and tuning the relative rotations and stacking orders amongst the crystalline layers will enable the development of an innumerable number of novel device characteristics.  This ability to combine two-dimensional materials into heterostructures is already showing promise for the creation of novel structures with tailored properties such as tunneling transistors and photovoltaic devices~\cite{Haigh:2012dx,Geim:2013dq,Georgiou:2012cd,Britnell:2013ca}.  Furthermore, with advances in the direct growth of these layers on top of each other~\cite{Yang:2013ev,Ding:2011eo,Son:2011gk,Tang:2012ju,Shi:2012dg}, there is great hope for industry scalable devices utilizing the incredible properties discovered in laboratory settings.  Local probe measurements have and will be an invaluable tool for understanding the local physics of these heterostructures.  

\section{References}

\bibliographystyle{unsrt}
\bibliography{references}

\begin{thebibliography}{100}

\bibitem{Novoselov:2004ub}
K~S Novoselov, A~K Geim, S~V Morozov, D~Jiang, Y~Zhang, S~V Dubonos, I~V
  Grigorieva, and A~A Firsov.
\newblock {Electric field effect in atomically thin carbon films}.
\newblock {\em Science (New York, NY)}, 306(5696):666--669, October 2004.

\bibitem{dasSarma:2011br}
S~das Sarma, Shaffique Adam, E~Hwang, and Enrico Rossi.
\newblock {Electronic transport in two-dimensional graphene}.
\newblock {\em Reviews of Modern Physics}, 83(2):407--470, May 2011.

\bibitem{Zhou:2006bt}
S~Y Zhou, G~H Gweon, J~Graf, A~V Fedorov, C~D Spataru, R~D Diehl, Y~Kopelevich,
  D~H Lee, Steven~G Louie, and A~Lanzara.
\newblock {First direct observation of Dirac fermions in graphite}.
\newblock {\em Nature Physics}, 2(9):595--599, September 2006.

\bibitem{Ferrari:2007fb}
Andrea Ferrari.
\newblock {Raman spectroscopy of graphene and graphite: Disorder,
  electron--phonon coupling, doping and nonadiabatic effects}.
\newblock {\em Solid State Communications}, 143(1-2):47--57, July 2007.

\bibitem{Morgenstern:2011hh}
Markus Morgenstern.
\newblock {Scanning tunneling microscopy and spectroscopy of graphene on
  insulating substrates}.
\newblock {\em Physica Status Solidi B-Basic Solid State Physics},
  248(11):2423--2434, November 2011.

\bibitem{Deshpande:2012jg}
Aparna Deshpande and Brian~J LeRoy.
\newblock {Scanning probe microscopy of graphene}.
\newblock {\em Physica E: Low-dimensional Systems and Nanostructures},
  44(4):743--759, January 2012.

\bibitem{Andrei:2012hy}
Eva~Y Andrei, Guohong Li, and Xu~Du.
\newblock {Electronic properties of graphene: a perspective from scanning
  tunneling microscopy and magnetotransport}.
\newblock {\em Reports on Progress in Physics}, 75(5):056501, April 2012.

\bibitem{Dean:2010jy}
C~R Dean, A~F Young, I~Meric, C~Lee, L~Wang, S~Sorgenfrei, K~Watanabe,
  T~Taniguchi, P~Kim, K~L Shepard, and J~Hone.
\newblock {Boron nitride substrates for high-quality graphene electronics}.
\newblock {\em Nature Nanotechnology}, 5(10):722--726, October 2010.

\bibitem{Dean:2012ht}
C~Dean, A~F Young, L~Wang, I~Meric, G.-H. Lee, K~Watanabe, T~Taniguchi,
  K~Shepard, P~Kim, and J~Hone.
\newblock {Graphene based heterostructures}.
\newblock {\em Solid State Communications}, 152(15):1275--1282, August 2012.

\bibitem{Gannett:2011gs}
W.~Gannett, W.~Regan, K~Watanabe, T~Taniguchi, M.~F. Crommie, and A.~Zettl.
\newblock {Boron nitride substrates for high mobility chemical vapor deposited
  graphene}.
\newblock {\em Applied Physics Letters}, 98(24):242105, June 2011.

\bibitem{Kim:2011ju}
Edwin Kim, Tianhua Yu, Eui Sang~Song, and Bin Yu.
\newblock {Chemical vapor deposition-assembled graphene field-effect transistor
  on hexagonal boron nitride}.
\newblock {\em Applied Physics Letters}, 98(26):262103, June 2011.

\bibitem{Taychatanapat:2011hr}
Thiti Taychatanapat, Kenji Watanabe, Takashi Taniguchi, and Pablo
  Jarillo-Herrero.
\newblock {Quantum Hall effect and Landau-level crossing of Dirac fermions in
  trilayer graphene}.
\newblock {\em Nature Physics}, 7(8):621--625, August 2011.

\bibitem{Zomer:2011ft}
P.~J. Zomer, S.~P. Dash, N.~Tombros, and B.~J. van Wees.
\newblock {A transfer technique for high mobility graphene devices on
  commercially available hexagonal boron nitride}.
\newblock {\em Applied Physics Letters}, 99(23):232104, December 2011.

\bibitem{SanchezYamagishi:2012jv}
Javier~D Sanchez-Yamagishi, Thiti Taychatanapat, Kenji Watanabe, Takashi
  Taniguchi, Amir Yacoby, and Pablo Jarillo-Herrero.
\newblock {Quantum Hall Effect, Screening, and Layer-Polarized Insulating
  States in Twisted Bilayer Graphene}.
\newblock {\em Physical Review Letters}, 108(7):076601, February 2012.

\bibitem{Young:2012bn}
A~F Young, C~R Dean, L~Wang, H~Ren, P~Cadden-Zimansky, K~Watanabe, T~Taniguchi,
  J~Hone, K~L Shepard, and P~Kim.
\newblock {Spin and valley quantum Hall ferromagnetism in graphene}.
\newblock {\em Nature Physics}, 8(7):550--556, July 2012.

\bibitem{Lee:2014tn}
Kayoung Lee, B~Fallahazad, J~Xue, T~Taniguchi, K~Watanabe, and E~Tutuc.
\newblock {Chemical potential and quantum Hall ferromagnetism in bilayer
  graphene mapped using double bilayer heterostructures}.
\newblock {\em arXiv}, page 1401.0659, January 2014.

\bibitem{Dean:2011ks}
C~R Dean, A~F Young, P~Cadden-Zimansky, L~Wang, H~Ren, K~Watanabe, T~Taniguchi,
  P~Kim, J~Hone, and K~L Shepard.
\newblock {Multicomponent fractional quantum Hall effect in graphene}.
\newblock {\em Nature Physics}, 7(9):693--696, September 2011.

\bibitem{Young:2013ft}
A~F Young, J~D Sanchez-Yamagishi, B~Hunt, S~H Choi, K~Watanabe, T~Taniguchi,
  R~C Ashoori, and P~Jarillo-Herrero.
\newblock {Tunable symmetry breaking and helical edge transport in a graphene
  quantum spin Hall state}.
\newblock {\em Nature}, pages~--, December 2013.

\bibitem{Amet:2013tk}
F~Amet, J~R Wiliams, K~Watanabe, T~Taniguchi, and D~Goldhaber-Gordon.
\newblock {Selective equilibration of spin and valley polarized quantum Hall
  edge states in graphene}.
\newblock {\em arXiv}, page 1307.4408, July 2013.

\bibitem{Ponomarenko:2011cj}
L~A Ponomarenko, A~K Geim, A~A Zhukov, R~Jalil, S~V Morozov, K~S Novoselov, I~V
  Grigorieva, E~H Hill, V~V Cheianov, V~I Fal'ko, K~Watanabe, T~Taniguchi, and
  R~V Gorbachev.
\newblock {Tunable metal-insulator transition in double-layer graphene
  heterostructures}.
\newblock {\em Nature Physics}, 7(12):958--961, December 2011.

\bibitem{Hunt:2013ef}
B~Hunt, J~D Sanchez-Yamagishi, A~F Young, M~Yankowitz, Brian~J LeRoy,
  K~Watanabe, T~Taniguchi, P~Moon, M~Koshino, P~Jarillo-Herrero, and R~C
  Ashoori.
\newblock {Massive Dirac Fermions and Hofstadter Butterfly in a van der Waals
  Heterostructure}.
\newblock {\em Science (New York, NY)}, 340(6139):1427--1430, May 2013.

\bibitem{Amet:2013gw}
F~Amet, J~R Williams, K~Watanabe, T~Taniguchi, and D~Goldhaber-Gordon.
\newblock {Insulating Behavior at the Neutrality Point in Single-Layer
  Graphene}.
\newblock {\em Physical Review Letters}, 110(21):216601, May 2013.

\bibitem{Woods:2014co}
C~R Woods, L~Britnell, A~Eckmann, R~S Ma, J~C Lu, H~M Guo, X~Lin, G~L Yu,
  Y~Cao, R~V Gorbachev, A~V Kretinin, J~Park, L~A Ponomarenko, M~I Katsnelson,
  Yu~N Gornostyrev, K~Watanabe, T~Taniguchi, C~Casiraghi, H~J Gao, A~K Geim,
  and K~S Novoselov.
\newblock {Commensurate--incommensurate transition in graphene on hexagonal
  boron nitride}.
\newblock {\em Nature Physics}, advance online publication 28 April 2014, DOI:
  10.1038/nphys2954.

\bibitem{Gorbachev:2012bn}
R~V Gorbachev, A~K Geim, M~I Katsnelson, K~S Novoselov, T~Tudorovskiy, I~V
  Grigorieva, A~H Macdonald, S~V Morozov, K~Watanabe, T~Taniguchi, and L~A
  Ponomarenko.
\newblock {Strong Coulomb drag and broken symmetry in double-layer graphene}.
\newblock {\em Nature Physics}, 8(12):896--901, December 2012.

\bibitem{Titov:2013gg}
M~Titov, R~V Gorbachev, B~N Narozhny, T~Tudorovskiy, M~Sch{\"u}tt, P~M
  Ostrovsky, I~V Gornyi, A~D Mirlin, M~I Katsnelson, K~S Novoselov, A~K Geim,
  and L~A Ponomarenko.
\newblock {Giant Magnetodrag in Graphene at Charge Neutrality}.
\newblock {\em Physical Review Letters}, 111(16):166601, October 2013.

\bibitem{Ponomarenko:2013hl}
L~A Ponomarenko, R~V Gorbachev, G~L Yu, D~C Elias, R~Jalil, A~A Patel,
  A~Mishchenko, A~S Mayorov, C~R Woods, J~R Wallbank, M~Mucha-Kruczynski, B~A
  Piot, M~Potemski, I~V Grigorieva, K~S Novoselov, F~Guinea, V~I Fal'ko, and
  A~K Geim.
\newblock {Cloning of Dirac fermions in graphene superlattices}.
\newblock {\em Nature}, 497(7451):594--597, May 2013.

\bibitem{Dean:2013bv}
C~R Dean, L~Wang, P~Maher, C~Forsythe, F~Ghahari, Y~Gao, J~Katoch, M~Ishigami,
  P~Moon, M~Koshino, T~Taniguchi, K~Watanabe, K~L Shepard, J~Hone, and P~Kim.
\newblock {Hofstadter's butterfly and the fractal quantum Hall effect in
  moir{\'e} superlattices}.
\newblock {\em Nature}, 497(7451):598--602, May 2013.

\bibitem{Taychatanapat:2013ek}
Thiti Taychatanapat, Kenji Watanabe, Takashi Taniguchi, and Pablo
  Jarillo-Herrero.
\newblock {Electrically tunable transverse magnetic focusing in graphene}.
\newblock {\em Nature Physics}, 9(4):225--229, April 2013.

\bibitem{Abanin:2011jq}
D~A Abanin, S~V Morozov, L~A Ponomarenko, R~V Gorbachev, A~S Mayorov, M~I
  Katsnelson, K~Watanabe, T~Taniguchi, K~S Novoselov, L~S Levitov, and A~K
  Geim.
\newblock {Giant Nonlocality Near the Dirac Point in Graphene}.
\newblock {\em Science (New York, NY)}, 332(6027):328--330, April 2011.

\bibitem{Wang:2011je}
Han Wang, Thiti Taychatanapat, Allen Hsu, Kenji Watanabe, Takashi Taniguchi,
  Pablo Jarillo-Herrero, and Tomas Palacios.
\newblock {BN/Graphene/BN Transistors for RF Applications}.
\newblock {\em IEEE Electron Device Letters}, 32(9):1209--1211, September 2011.

\bibitem{Amet:2012eu}
F~Amet, J~R Williams, A~G~F Garcia, M~Yankowitz, K~Watanabe, T~Taniguchi, and
  D~Goldhaber-Gordon.
\newblock {Tunneling spectroscopy of graphene-boron-nitride heterostructures}.
\newblock {\em Physical Review B}, 85(7):073405, February 2012.

\bibitem{Britnell:2012jp}
L~Britnell, R~V Gorbachev, R~Jalil, B~D Belle, F~Schedin, A~Mishchenko,
  T~Georgiou, M~I Katsnelson, L~Eaves, S~V Morozov, N~M~R Peres, J~Leist, A~K
  Geim, K~S Novoselov, and L~A Ponomarenko.
\newblock {Field-Effect Tunneling Transistor Based on Vertical Graphene
  Heterostructures}.
\newblock {\em Science (New York, NY)}, 335(6071):947--950, February 2012.

\bibitem{Britnell:2012dq}
Liam Britnell, Roman~V Gorbachev, Rashid Jalil, Branson~D Belle, Fred Schedin,
  Mikhail~I. Katsnelson, Laurence Eaves, Sergey~V Morozov, Alexander~S Mayorov,
  Nuno M~R Peres, Antonio~H Castro~Neto, Jon Leist, Andre~K. Geim, Leonid~A
  Ponomarenko, and Kostya~S Novoselov.
\newblock {Electron Tunneling through Ultrathin Boron Nitride Crystalline
  Barriers}.
\newblock {\em Nano Letters}, 12(3):1707--1710, March 2012.

\bibitem{Young:2012dv}
A~F Young, C~R Dean, I~Meric, S~Sorgenfrei, H~Ren, K~Watanabe, T~Taniguchi,
  J~Hone, K~L Shepard, and P~Kim.
\newblock {Electronic compressibility of layer-polarized bilayer graphene}.
\newblock {\em Physical Review B}, 85(23):235458, June 2012.

\bibitem{Masubuchi:2012cz}
Satoru Masubuchi, Kazuyuki Iguchi, Takehiro Yamaguchi, Masahiro Onuki, Miho
  Arai, Kenji Watanabe, Takashi Taniguchi, and Tomoki Machida.
\newblock {Boundary Scattering in Ballistic Graphene}.
\newblock {\em Physical Review Letters}, 109(3):036601, July 2012.

\bibitem{Haigh:2012dx}
S~J Haigh, A~Gholinia, R~Jalil, S~Romani, L~Britnell, D~C Elias, K~S Novoselov,
  L~A Ponomarenko, A~K Geim, and R~Gorbachev.
\newblock {Cross-sectional imaging of individual layers and buried interfaces
  of graphene-based heterostructures and superlattices}.
\newblock {\em Nature Materials}, 11(9):764--767, September 2012.

\bibitem{Sutar:2012br}
S~Sutar, E~S Comfort, J~Liu, T~Taniguchi, K~Watanabe, and J~U Lee.
\newblock {Angle-Dependent Carrier Transmission in Graphene p--n Junctions}.
\newblock {\em Nano Letters}, 12(9):4460--4464, September 2012.

\bibitem{Goossens:2012jq}
Augustinus Stijn~M Goossens, Stefanie C~M Driessen, Tim~A Baart, Kenji
  Watanabe, Takashi Taniguchi, and Lieven M~K Vandersypen.
\newblock {Gate-Defined Confinement in Bilayer Graphene-Hexagonal Boron Nitride
  Hybrid Devices}.
\newblock {\em Nano Letters}, 12(9):4656--4660, September 2012.

\bibitem{Droscher:2012ir}
Susanne Dr{\"o}scher, Cl{\'e}ment Barraud, Kenji Watanabe, Takashi Taniguchi,
  Thomas Ihn, and Klaus Ensslin.
\newblock {Electron flow in split-gated bilayer graphene}.
\newblock {\em New Journal of Physics}, 14(10):103007, October 2012.

\bibitem{Zomer:2012gw}
P.~J. Zomer, M~H~D Guimar{\~a}es, N.~Tombros, and B.~J. van Wees.
\newblock {Long-distance spin transport in high-mobility graphene on hexagonal
  boron nitride}.
\newblock {\em Physical Review B}, 86(16):161416, October 2012.

\bibitem{Bischoff:2012hf}
D~Bischoff, T~Kr{\"a}henmann, S~Dr{\"o}scher, M~A Gruner, C~Barraud, T~Ihn, and
  K~Ensslin.
\newblock {Reactive-ion-etched graphene nanoribbons on a hexagonal boron
  nitride substrate}.
\newblock {\em Applied Physics Letters}, 101(20):203103, November 2012.

\bibitem{Campos:2012je}
L~C Campos, A~F Young, K~Surakitbovorn, K~Watanabe, T~Taniguchi, and
  P~Jarillo-Herrero.
\newblock {Quantum and classical confinement of resonant states in a trilayer
  graphene Fabry-P{\'e}rot interferometer}.
\newblock {\em Nature Communications}, 3:1239, December 2012.

\bibitem{Yu:2013ku}
G~L Yu, R~Jalil, Branson Belle, Alexander~S Mayorov, Peter Blake, Frederick
  Schedin, Sergey~V Morozov, Leonid~A Ponomarenko, F~Chiappini, S~Wiedmann, Uli
  Zeitler, Mikhail~I. Katsnelson, A~K Geim, Kostya~S Novoselov, and Daniel~C
  Elias.
\newblock {Interaction phenomena in graphene seen through quantum capacitance}.
\newblock {\em Proceedings of the National Academy of Sciences},
  110(9):3282--3286, February 2013.

\bibitem{Maher:2013gw}
P~Maher, C~R Dean, A~F Young, T~Taniguchi, K~Watanabe, K~L Shepard, J~Hone, and
  P~Kim.
\newblock {Evidence for a spin phase transition at charge neutrality in bilayer
  graphene}.
\newblock {\em Nature Physics}, 9(3):154--158, March 2013.

\bibitem{Ponomarenko:2013dr}
L~A Ponomarenko, B~D Belle, R~Jalil, L~Britnell, R~V Gorbachev, A~K Geim, K~S
  Novoselov, A~H Castro~Neto, L~Eaves, and M~I Katsnelson.
\newblock {Field-effect control of tunneling barrier height by exploiting
  graphene's low density of states}.
\newblock {\em Journal of Applied Physics}, 113(13):136502, March 2013.

\bibitem{Britnell:2013ku}
L~Britnell, R~V Gorbachev, A~K Geim, L~A Ponomarenko, A~Mishchenko, M~T
  Greenaway, T~M Fromhold, K~S Novoselov, and L~Eaves.
\newblock {Resonant tunnelling and negative differential conductance in
  graphene transistors}.
\newblock {\em Nature Communications}, 4:1794, April 2013.

\bibitem{Meric:2011tm}
Inanc Meric, Cory~R Dean, Nicholas Petrone, Lei Wang, James Hone, Philip Kim,
  and Kenneth~L Shepard.
\newblock {Graphene Field-Effect Transistors Based on Boron-Nitride
  Dielectrics}.
\newblock {\em Proceedings Of The Ieee}, 101(7):1609--1619, July 2013.

\bibitem{Engels:2013fu}
S~Engels, A~Epping, C~Volk, S~Korte, B~Voigtl{\"a}nder, K~Watanabe,
  T~Taniguchi, S~Trellenkamp, and C~Stampfer.
\newblock {Etched graphene quantum dots on hexagonal boron nitride}.
\newblock {\em Applied Physics Letters}, 103(7):073113, August 2013.

\bibitem{Wang:2013ch}
L~Wang, I~Meric, P~Y Huang, Q~Gao, Y~Gao, H~Tran, T~Taniguchi, K~Watanabe, L~M
  Campos, D~A Muller, J~Guo, P~Kim, J~Hone, K~L Shepard, and C~R Dean.
\newblock {One-Dimensional Electrical Contact to a Two-Dimensional Material}.
\newblock {\em Science (New York, NY)}, 342(6158):614--617, October 2013.

\bibitem{Epping:2013kd}
A~Epping, S~Engels, C~Volk, K~Watanabe, T~Taniguchi, S~Trellenkamp, and
  C~Stampfer.
\newblock {Etched graphene single electron transistors on hexagonal boron
  nitride in high magnetic fields}.
\newblock {\em physica status solidi (b)}, 250(12):2692--2696, November 2013.

\bibitem{Wang:2012ex}
Lei Wang, Zheyuan Chen, Cory~R Dean, Takashi Taniguchi, Kenji Watanabe, Louis~E
  Brus, and James Hone.
\newblock {Negligible Environmental Sensitivity of Graphene in a Hexagonal
  Boron Nitride/Graphene/h-BN Sandwich Structure}.
\newblock {\em ACS Nano}, page 121001151424005, October 2012.

\bibitem{Ahn:2013fa}
Gwanghyun Ahn, Hye~Ri Kim, Taeg~Yeoung Ko, Kyoungjun Choi, Kenji Watanabe,
  Takashi Taniguchi, Byung~Hee Hong, and Sunmin Ryu.
\newblock {Optical Probing of the Electronic Interaction between Graphene and
  Hexagonal Boron Nitride}.
\newblock {\em ACS Nano}, 7(2):1533--1541, February 2013.

\bibitem{Forster:2013fi}
F~Forster, A~Molina-Sanchez, S~Engels, A~Epping, K~Watanabe, T~Taniguchi,
  L~Wirtz, and C~Stampfer.
\newblock {Dielectric screening of the Kohn anomaly of graphene on hexagonal
  boron nitride}.
\newblock {\em Physical Review B}, 88(8):085419, August 2013.

\bibitem{Chattrakun:2013bt}
Kanokporn Chattrakun, Shengqiang Huang, K~Watanabe, T~Taniguchi, A~Sandhu, and
  Brian~J LeRoy.
\newblock {Gate dependent Raman spectroscopy of graphene on hexagonal boron
  nitride}.
\newblock {\em Journal Of Physics-Condensed Matter}, 25(50):505304, November
  2013.

\bibitem{Abergel:2013el}
D~S~L Abergel, J~R Wallbank, Xi~Chen, M~Mucha-Kruczynski, and Vladimir~I
  Fal'ko.
\newblock {Infrared absorption by graphene--hBN heterostructures}.
\newblock {\em New Journal of Physics}, 15(12):123009, December 2013.

\bibitem{Ju:2014dz}
L~Ju, J~Velasco, E~Huang, S~Kahn, C~Nosiglia, Hsin-Zon Tsai, W~Yang,
  T~Taniguchi, K~Watanabe, Y~Zhang, G~Zhang, M~Crommie, A.~Zettl, and F~Wang.
\newblock {Photoinduced doping in heterostructures of graphene and boron
  nitride}.
\newblock {\em Nature Nanotechnology}, advance online publication 13 April
  2014, DOI: 10.1038/nnano.2014.60.

\bibitem{Neto:2009cl}
A~H Castro~Neto, F~Guinea, N~M~R Peres, K~S Novoselov, and A~K Geim.
\newblock {The electronic properties of graphene}.
\newblock {\em Reviews of Modern Physics}, 81(1):109--162, January 2009.

\bibitem{McCann:2012dc}
Edward McCann.
\newblock {Electronic Properties of Monolayer and Bilayer Graphene}.
\newblock In Hassan Raza, editor, {\em Graphene Nanoelectronics}, pages
  237--275. Springer Berlin Heidelberg, Berlin, Heidelberg, 2012.

\bibitem{Katsnelson:2006kd}
M~I Katsnelson, K~S Novoselov, and A~K Geim.
\newblock {Chiral tunnelling and the Klein paradox in graphene}.
\newblock {\em Nature Physics}, 2(9):620--625, September 2006.

\bibitem{Stander:2009ce}
N~Stander, B~Huard, and D~Goldhaber-Gordon.
\newblock {Evidence for Klein Tunneling in Graphene p-n Junctions}.
\newblock {\em Physical Review Letters}, 102(2):026807, January 2009.

\bibitem{Young:2009il}
Andrea~F Young and Philip Kim.
\newblock {Quantum interference and Klein tunnelling in graphene
  heterojunctions}.
\newblock {\em Nature Physics}, 5(3):222--226, March 2009.

\bibitem{Park:2008eg}
Cheol-Hwan Park, Li~Yang, Young-Woo Son, Marvin~L Cohen, and Steven~G Louie.
\newblock {Anisotropic behaviours of massless Dirac fermions in graphene under
  periodic potentials}.
\newblock {\em Nature Physics}, 4(3):213--217, March 2008.

\bibitem{Park:2008kia}
Cheol-Hwan Park, Li~Yang, Young-Woo Son, Marvin~L Cohen, and Steven~G Louie.
\newblock {New Generation of Massless Dirac Fermions in Graphene under External
  Periodic Potentials}.
\newblock {\em Physical Review Letters}, 101(12):126804, September 2008.

\bibitem{Barbier:2008gx}
Micha{\"e}l Barbier, F~Peeters, P~Vasilopoulos, and J~Pereira.
\newblock {Dirac and Klein-Gordon particles in one-dimensional periodic
  potentials}.
\newblock {\em Physical Review B}, 77(11):115446, March 2008.

\bibitem{Brey:2009jx}
L~Brey and H~Fertig.
\newblock {Emerging Zero Modes for Graphene in a Periodic Potential}.
\newblock {\em Physical Review Letters}, 103(4):046809, July 2009.

\bibitem{Barbier:2009ks}
M~Barbier, P~Vasilopoulos, and F~M Peeters.
\newblock {Dirac electrons in a Kronig-Penney potential: Dispersion relation
  and transmission periodic in the strength of the barriers}.
\newblock {\em Physical Review B}, 80(20):205415, November 2009.

\bibitem{Sun:2010fx}
Jianmin Sun, H~A Fertig, and L~Brey.
\newblock {Effective Magnetic Fields in Graphene Superlattices}.
\newblock {\em Physical Review Letters}, 105(15):156801, October 2010.

\bibitem{Burset:2011cc}
P~Burset, A~Yeyati, L~Brey, and H~Fertig.
\newblock {Transport in superlattices on single-layer graphene}.
\newblock {\em Physical Review B}, 83(19):195434, May 2011.

\bibitem{Ortix:2012tm}
C~Ortix, L~Yang, and J~van~den Brink.
\newblock {Graphene on incommensurate substrates: Trigonal warping and emerging
  Dirac cone replicas with halved group velocity}.
\newblock {\em Physical Review B}, 86(8):081405, August 2012.

\bibitem{Wallbank:2013kz}
J~R Wallbank, M~Mucha-Kruczynski, and V~I Fal'ko.
\newblock {Moire minibands in graphene heterostructures with almost
  commensurate $\sqrt{3} x \sqrt{3}$ hexagonal crystals}.
\newblock {\em Physical Review B}, 88(15):155415, October 2013.

\bibitem{Wallbank:2013ep}
J~R Wallbank, A~A Patel, M~Mucha-Kruczynski, A~K Geim, and V~I Fal'ko.
\newblock {Generic miniband structure of graphene on a hexagonal substrate}.
\newblock {\em Physical Review B}, 87(24):245408, June 2013.

\bibitem{MuchaKruczynski:2013cw}
M~Mucha-Kruczynski, J~R Wallbank, and V~I Fal'ko.
\newblock {Heterostructures of bilayer graphene and h-BN: Interplay between
  misalignment, interlayer asymmetry, and trigonal warping}.
\newblock {\em Physical Review B}, 88(20):205418, November 2013.

\bibitem{Jung:2013tq}
Jeil Jung, Arnaud Raoux, Zhenhua Qiao, and Allan~H MacDonald.
\newblock {Ab-Initio Theory of Moir\'e Superlattice Bands in Layered
  Two-Dimensional Materials}.
\newblock {\em arXiv}, page 1312.7723, December 2013.

\bibitem{Chen:2014ek}
Xi~Chen, J~R Wallbank, A~A Patel, M~Mucha-Kruczynski, E~Mccann, and V~I Fal'ko.
\newblock {Dirac edges of fractal magnetic minibands in graphene with hexagonal
  moir{\'e} superlattices}.
\newblock {\em Physical Review B}, 89(7):075401, February 2014.

\bibitem{Yankowitz:2012gi}
Matthew Yankowitz, Jiamin Xue, Daniel Cormode, Javier~D Sanchez-Yamagishi,
  K~Watanabe, T~Taniguchi, Pablo Jarillo-Herrero, Philippe Jacquod, and Brian~J
  LeRoy.
\newblock {Emergence of superlattice Dirac points in graphene on hexagonal
  boron nitride}.
\newblock {\em Nature Physics}, 8(5):382--386, May 2012.

\bibitem{Watanabe:2004ct}
Kenji Watanabe, Takashi Taniguchi, and Hisao Kanda.
\newblock {Direct-bandgap properties and evidence for ultraviolet lasing of
  hexagonal boron nitride single crystal}.
\newblock {\em Nature Materials}, 3(6):404--409, June 2004.

\bibitem{Chen:2008vn}
C~Julian Chen.
\newblock {\em {Introduction to scanning tunneling microscopy}}.
\newblock Oxford University Press, USA, 2nd edition, 2008.

\bibitem{Crommie:1993vk}
Michael Crommie, Chris Lutz, and Don Eigler.
\newblock {Confinement of electrons to quantum corrals on a metal-surface}.
\newblock {\em Science (New York, NY)}, 262(5131):218--220, October 1993.

\bibitem{Cullen:2010ib}
W~Cullen, M~Yamamoto, K~Burson, J~Chen, C~Jang, L~Li, M~Fuhrer, and E~Williams.
\newblock {High-Fidelity Conformation of Graphene to SiO$_{2}$ Topographic
  Features}.
\newblock {\em Physical Review Letters}, 105(21):215504, November 2010.

\bibitem{Lui:2009kv}
Chun~Hung Lui, Li~Liu, Kin~Fai Mak, George~W Flynn, and Tony~F Heinz.
\newblock {Ultraflat graphene}.
\newblock {\em Nature}, 462(7271):339--341, November 2009.

\bibitem{Xue:2011dv}
Jiamin Xue, Javier Sanchez-Yamagishi, Danny Bulmash, Philippe Jacquod, Aparna
  Deshpande, K~Watanabe, T~Taniguchi, Pablo Jarillo-Herrero, and Brian~J LeRoy.
\newblock {Scanning tunnelling microscopy and spectroscopy of ultra-flat
  graphene on hexagonal boron nitride}.
\newblock {\em Nature Materials}, 10(4):282--285, April 2011.

\bibitem{Decker:2011wh}
Regis Decker, Yang Wang, Victor~W Brar, William Regan, Hsin-Zon Tsai, Qiong Wu,
  William Gannett, Alex Zettl, and Michael~F Crommie.
\newblock {Local Electronic Properties of Graphene on a BN Substrate via
  Scanning Tunneling Microscopy}.
\newblock {\em Nano Letters}, 11(6):2291--2295, June 2011.

\bibitem{Roth:2013ca}
Silvan Roth, Fumihiko Matsui, Thomas Greber, and Juerg Osterwalder.
\newblock {Chemical Vapor Deposition and Characterization of Aligned and
  Incommensurate Graphene/Hexagonal Boron Nitride Heterostack on Cu(111)}.
\newblock {\em Nano Letters}, 13(6):2668--2675, June 2013.

\bibitem{Yang:2013ev}
W~Yang, G~Chen, Z~Shi, C~C Liu, L~Zhang, and G~Xie.
\newblock {Epitaxial growth of single-domain graphene on hexagonal boron
  nitride}.
\newblock {\em Nature Materials}, 12(9):792--797, September 2013.

\bibitem{Tang:2013hy}
Shujie Tang, Haomin Wang, Yu~Zhang, Ang Li, Hong Xie, Xiaoyu Liu, Lianqing Liu,
  Tianxin Li, Fuqiang Huang, Xiaoming Xie, and Mianheng Jiang.
\newblock {Precisely aligned graphene grown on hexagonal boron nitride by
  catalyst free chemical vapor deposition}.
\newblock {\em Scientific Reports}, 3:2666, September 2013.

\bibitem{Wintterlin:2009bc}
J~Wintterlin and M-L Bocquet.
\newblock {Graphene on metal surfaces}.
\newblock {\em Surface Science}, 603(10-12):1841--1852, January 2009.

\bibitem{Li:2009eh}
Guohong Li, Adina Luican, and Eva~Y Andrei.
\newblock {Scanning Tunneling Spectroscopy of Graphene on Graphite}.
\newblock {\em Physical Review Letters}, 102(17):176804, May 2009.

\bibitem{Rutter:2007epa}
G~M Rutter, J~N Crain, N~P Guisinger, T~Li, P~N First, and J~A Stroscio.
\newblock {Scattering and Interference in Epitaxial Graphene}.
\newblock {\em Science (New York, NY)}, 317(5835):219--222, July 2007.

\bibitem{Gao:2010iz}
Li~Gao, Jeffrey~R Guest, and Nathan~P Guisinger.
\newblock {Epitaxial Graphene on Cu(111)}.
\newblock {\em Nano Letters}, 10(9):3512--3516, September 2010.

\bibitem{Ndiaye:2006cr}
Alpha~T N'Diaye, Sebastian Bleikamp, Peter~J Feibelman, and Thomas Michely.
\newblock {Two-dimensional Ir cluster lattice on a graphene moire on Ir(111)}.
\newblock {\em Physical Review Letters}, 97(21):215501, November 2006.

\bibitem{Dedkov:2010jh}
Yu~S Dedkov and M~Fonin.
\newblock {Electronic and magnetic properties of the graphene--ferromagnet
  interface}.
\newblock {\em New Journal of Physics}, 12(12):125004, December 2010.

\bibitem{Kwon:2009kn}
Soon-Yong Kwon, Cristian~V Ciobanu, Vania Petrova, Vivek~B Shenoy, Javier
  Bare{\~n}o, Vincent Gambin, Ivan Petrov, and Suneel Kodambaka.
\newblock {Growth of Semiconducting Graphene on Palladium}.
\newblock {\em Nano Letters}, 9(12):3985--3990, December 2009.

\bibitem{Sutter:2009ff}
Peter Sutter, Jerzy~T Sadowski, and Eli Sutter.
\newblock {Graphene on Pt(111): Growth and substrate interaction}.
\newblock {\em Physical Review B}, 80(24):245411, December 2009.

\bibitem{Roth:2011ja}
Silvan Roth, Juerg Osterwalder, and Thomas Greber.
\newblock {Synthesis of epitaxial graphene on rhodium from 3-pentanone}.
\newblock {\em Surface Science}, 605(9-10):L17--L19, May 2011.

\bibitem{Marchini:2007gt}
S~Marchini, S~G{\"u}nther, and J~Wintterlin.
\newblock {Scanning tunneling microscopy of graphene on Ru(0001)}.
\newblock {\em Physical Review B}, 76(7):075429, August 2007.

\bibitem{Giovannetti:2007jc}
Gianluca Giovannetti, Petr Khomyakov, Geert Brocks, Paul Kelly, and Jeroen
  van~den Brink.
\newblock {Substrate-induced band gap in graphene on hexagonal boron nitride:
  Ab initio density functional calculations}.
\newblock {\em Physical Review B}, 76(7):073103, August 2007.

\bibitem{Siawinska:2010kr}
J~S{\l}awi{\'n}ska, I~Zasada, and Z~Klusek.
\newblock {Energy gap tuning in graphene on hexagonal boron nitride bilayer
  system}.
\newblock {\em Physical Review B}, 81(15):155433, April 2010.

\bibitem{Sachs:2011jp}
B.~Sachs, T.~O. Wehling, M~I Katsnelson, and A.~I. Lichtenstein.
\newblock {Adhesion and electronic structure of graphene on hexagonal boron
  nitride substrates}.
\newblock {\em Physical Review B}, 84(19):195414, November 2011.

\bibitem{Kindermann:2012jz}
M~Kindermann, Bruno Uchoa, and D~L Miller.
\newblock {Zero-energy modes and gate-tunable gap in graphene on hexagonal
  boron nitride}.
\newblock {\em Physical Review B}, 86(11):115415, September 2012.

\bibitem{Martin:2008ca}
Jens Martin, N~Akerman, G~Ulbricht, T~Lohmann, J~H Smet, K~von Klitzing, and
  Amir Yacoby.
\newblock {Observation of electron-hole puddles in graphene using a scanning
  single-electron transistor}.
\newblock {\em Nature Physics}, 4(2):144--148, February 2008.

\bibitem{Deshpande:2009hx}
Aparna Deshpande, Wenzhong Bao, F~Miao, Chun~Ning Lau, and Brian~J LeRoy.
\newblock {Spatially resolved spectroscopy of monolayer graphene on SiO$_2$}.
\newblock {\em Physical Review B}, 79(20):205411, May 2009.

\bibitem{Zhang:2009ce}
Yuanbo Zhang, Victor~W Brar, Caglar Girit, Alex Zettl, and Michael~F Crommie.
\newblock {Origin of spatial charge inhomogeneity in graphene}.
\newblock {\em Nature Physics}, 5(10):722--726, October 2009.

\bibitem{Burson:2013fc}
Kristen~M Burson, William~G Cullen, Shaffique Adam, Cory~R Dean, K~Watanabe,
  T~Taniguchi, Philip Kim, and Michael~S Fuhrer.
\newblock {Direct Imaging of Charged Impurity Density in Common Graphene
  Substrates}.
\newblock {\em Nano Letters}, 13(8):3576--3580, August 2013.

\bibitem{Eckmann:2013da}
Axel Eckmann, Jaesung Park, Huafeng Yang, Daniel Elias, Alexander~S Mayorov,
  Geliang Yu, Rashid Jalil, Kostya~S Novoselov, Roman~V Gorbachev, Michele
  Lazzeri, Andre~K. Geim, and Cinzia Casiraghi.
\newblock {Raman Fingerprint of Aligned Graphene/h-BN Superlattices}.
\newblock {\em Nano Letters}, 13(11):5242--5246, November 2013.

\bibitem{NeekAmal:2014ig}
M~Neek-Amal and F~M Peeters.
\newblock {Graphene on boron-nitride: Moir{\'e} pattern in the van der Waals
  energy}.
\newblock {\em Applied Physics Letters}, 104(4):041909, January 2014.

\bibitem{Pletikosic:2009fh}
I~Pletikosi{\'c}, M~Kralj, P~Pervan, R~Brako, J~Coraux, A~N'Diaye, C~Busse, and
  T.~Michely.
\newblock {Dirac Cones and Minigaps for Graphene on Ir(111)}.
\newblock {\em Physical Review Letters}, 102(5):056808, February 2009.

\bibitem{Li:2009cp}
Guohong Li, A~Luican, J~M~B Lopes~dos Santos, A~H Castro~Neto, A~Reina, J~Kong,
  and E~Y Andrei.
\newblock {Observation of Van Hove singularities in twisted graphene layers}.
\newblock {\em Nature Physics}, 6(2):109--113, February 2010.

\bibitem{Luican:2011hw}
A~Luican, Guohong Li, A~Reina, J~Kong, R~Nair, K~Novoselov, A~Geim, and
  E~Andrei.
\newblock {Single-Layer Behavior and Its Breakdown in Twisted Graphene Layers}.
\newblock {\em Physical Review Letters}, 106(12):126802, March 2011.

\bibitem{Brihuega:2012hh}
I~Brihuega, P~Mallet, H~Gonz{\'a}lez-Herrero, G~Trambly~de Laissardi{\`e}re,
  M~Ugeda, L~Magaud, J~G{\'o}mez-Rodr{\'\i}guez, F~Yndur{\'a}in, and J-Y
  Veuillen.
\newblock {Unraveling the Intrinsic and Robust Nature of van Hove Singularities
  in Twisted Bilayer Graphene by Scanning Tunneling Microscopy and Theoretical
  Analysis}.
\newblock {\em Physical Review Letters}, 109(19):196802, November 2012.

\bibitem{Xue:2012hd}
Jiamin Xue, Javier Sanchez-Yamagishi, K~Watanabe, T~Taniguchi, Pablo
  Jarillo-Herrero, and Brian~J LeRoy.
\newblock {Long-Wavelength Local Density of States Oscillations Near Graphene
  Step Edges}.
\newblock {\em Physical Review Letters}, 108(1):016801, January 2012.

\bibitem{Crommie:1993co}
M.~F. Crommie, C~P Lutz, and D~M Eigler.
\newblock {Imaging standing waves in a two-dimensional electron gas}.
\newblock {\em Nature}, 363(6429):524--527, June 1993.

\bibitem{Hasegawa:1993dq}
Y~Hasegawa and Ph~Avouris.
\newblock {Direct observation of standing wave formation at surface steps using
  scanning tunneling spectroscopy}.
\newblock {\em Physical Review Letters}, 71(7):1071--1074, August 1993.

\bibitem{Ando:1998fn}
Tsuneya Ando, Takeshi Nakanishi, and Riichiro Saito.
\newblock {Berry's Phase and Absence of Back Scattering in Carbon Nanotubes}.
\newblock {\em Journal Of The Physical Society Of Japan}, 67(8):2857--2862,
  August 1998.

\bibitem{Zhou:2009fz}
Xiaoting Zhou, Chen Fang, Wei-Feng Tsai, and JiangPing Hu.
\newblock {Theory of quasiparticle scattering in a two-dimensional system of
  helical Dirac fermions: Surface band structure of a three-dimensional
  topological insulator}.
\newblock {\em Physical Review B}, 80(24):245317, December 2009.

\bibitem{Ohta:2006bo}
Taisuke Ohta, Aaron Bostwick, Thomas Seyller, Karsten Horn, and Eli Rotenberg.
\newblock {Controlling the electronic structure of bilayer graphene}.
\newblock {\em Science (New York, NY)}, 313(5789):951--954, August 2006.

\bibitem{Brar:2011iq}
Victor~W Brar, Regis Decker, Hans-Michael Solowan, Yang Wang, Lorenzo Maserati,
  Kevin~T Chan, Hoonkyung Lee, {\c C}a{\u g}lar~O Girit, Alex Zettl, Steven~G
  Louie, Marvin~L Cohen, and Michael~F Crommie.
\newblock {Gate-controlled ionization and screening of cobalt adatoms on a
  graphene surface}.
\newblock {\em Nature Physics}, 7(1):43--47, January 2011.

\bibitem{Wang:2012id}
Y~Wang, V~W Brar, A~V Shytov, Q~Wu, W.~Regan, H~Z Tsai, A.~Zettl, L~S Levitov,
  and M.~F. Crommie.
\newblock {Mapping Dirac quasiparticles near a single Coulomb impurity on
  graphene}.
\newblock {\em Nature Physics}, 8(9):653--657, September 2012.

\bibitem{Wang:2013ec}
Y~Wang, D~Wong, A~V Shytov, V~W Brar, S~Choi, Q~Wu, H~Z Tsai, W.~Regan,
  A.~Zettl, R~K Kawakami, S~G Louie, L~S Levitov, and M.~F. Crommie.
\newblock {Observing Atomic Collapse Resonances in Artificial Nuclei on
  Graphene}.
\newblock {\em Science (New York, NY)}, 340(6133):734--737, May 2013.

\bibitem{Darwin:1913gv}
C~G Darwin.
\newblock {On some orbits of an electron}.
\newblock {\em Philosophical Magazine Series 6}, 25(146):201--210, February
  1913.

\bibitem{Greiner:1985ij}
Walter Greiner, Berndt M{\"u}ller, and Johann Rafelski.
\newblock {\em {Quantum Electrodynamics of Strong Fields}}.
\newblock Springer Berlin Heidelberg, Berlin, Heidelberg, 1985.

\bibitem{Boyer:2004hu}
Timothy~H Boyer.
\newblock {Unfamiliar trajectories for a relativistic particle in a Kepler or
  Coulomb potential}.
\newblock {\em American Journal of Physics}, 72(8):992--997, August 2004.

\bibitem{Zeldovich:2007jk}
Ya~B Zeldovich and Valentin~S Popov.
\newblock {Electronic Structure of Superheavy Atoms}.
\newblock {\em Soviet Physics Uspekhi}, 14(6):673--694, October 2007.

\bibitem{Schweppe:1983fm}
J~Schweppe, A~Gruppe, K~Bethge, H~Bokemeyer, T~Cowan, H~Folger, J~Greenberg,
  H~Grein, S~Ito, R~Schule, D~Schwalm, K~Stiebing, N~Trautmann, P~Vincent, and
  M~Waldschmidt.
\newblock {Observation of a Peak Structure in Positron Spectra from U+Cm
  Collisions}.
\newblock {\em Physical Review Letters}, 51(25):2261--2264, December 1983.

\bibitem{Cowan:1985iu}
T~Cowan, H~Backe, M~Begemann, K~Bethge, H~Bokemeyer, H~Folger, J~Greenberg,
  H~Grein, A~Gruppe, Y~Kido, M~Kl{\"u}ver, D~Schwalm, J~Schweppe, K~Stiebing,
  N~Trautmann, and P~Vincent.
\newblock {Anomalous Positron Peaks from Supercritical Collision Systems}.
\newblock {\em Physical Review Letters}, 54(16):1761--1764, April 1985.

\bibitem{Pereira:2007di}
Vitor~M Pereira, Johan Nilsson, and A~H Castro~Neto.
\newblock {Coulomb Impurity Problem in Graphene}.
\newblock {\em Physical Review Letters}, 99(16):166802, October 2007.

\bibitem{Shytov:2007ck}
A~V Shytov, M~I Katsnelson, and L~S Levitov.
\newblock {Vacuum Polarization and Screening of Supercritical Impurities in
  Graphene}.
\newblock {\em Physical Review Letters}, 99(23):236801, December 2007.

\bibitem{Shytov:2007gu}
A~V Shytov, M~I Katsnelson, and L~S Levitov.
\newblock {Atomic Collapse and Quasi--Rydberg States in Graphene}.
\newblock {\em Physical Review Letters}, 99(24):246802, December 2007.

\bibitem{Chae:2012jk}
Jungseok Chae, Suyong Jung, Andrea~F Young, Cory~R Dean, Lei Wang, Yuanda Gao,
  Kenji Watanabe, Takashi Taniguchi, James Hone, Kenneth~L Shepard, Phillip
  Kim, Nikolai~B Zhitenev, and Joseph~A Stroscio.
\newblock {Renormalization of the Graphene Dispersion Velocity Determined from
  Scanning Tunneling Spectroscopy}.
\newblock {\em Physical Review Letters}, 109(11):116802, September 2012.

\bibitem{Zou:2013dx}
Q~Zou, B~D Belle, L~Z Zhang, W~D Xiao, K~Yang, L~W Liu, G~Q Wang, X~M Fei,
  Y~Huang, R~S Ma, Y~Lu, P~H Tan, H~M Guo, S~X Du, and H~J Gao.
\newblock {Modulation of Fermi velocities of Dirac electrons in single layer
  graphene by moir{\'e} superlattice}.
\newblock {\em Applied Physics Letters}, 103(11):113106, September 2013.

\bibitem{LuicanMayer:2014jt}
Adina Luican-Mayer, Maxim Kharitonov, Guohong Li, Chih-Pin Lu, Ivan Skachko,
  Alem-Mar~B Goncalves, K~Watanabe, T~Taniguchi, and Eva~Y Andrei.
\newblock {Screening Charged Impurities and Lifting the Orbital Degeneracy in
  Graphene by Populating Landau Levels}.
\newblock {\em Physical Review Letters}, 112(3):036804, January 2014.

\bibitem{Kou:2013ud}
A~Kou, B~E Feldman, Andrei~J Levin, Bertrand~I Halperin, Kenji Watanabe,
  Takashi Taniguchi, and Amir Yacoby.
\newblock {Electron-Hole Asymmetric Integer and Fractional Quantum Hall Effect
  in Bilayer Graphene}.
\newblock {\em arXiv}, page 1312.7033, December 2013.

\bibitem{Elias:2011ev}
D~C Elias, R~V Gorbachev, A~S Mayorov, S~V Morozov, A~A Zhukov, P~Blake, L~A
  Ponomarenko, I~V Grigorieva, K~S Novoselov, F~Guinea, and A~K Geim.
\newblock {Dirac cones reshaped by interaction effects in suspended graphene}.
\newblock {\em Nature Physics}, 7(9):701--704, September 2011.

\bibitem{DasSarma:1997hr}
Sankar Das~Sarma and Aron Pinczuk, editors.
\newblock {\em {Perspectives in Quantum Hall Effects}}.
\newblock Wiley-VCH Verlag GmbH, Weinheim, Germany, 1997.

\bibitem{Novoselov:2005es}
K~S Novoselov, A~K Geim, S~V Morozov, D~Jiang, M~I Katsnelson, I~V Grigorieva,
  S~V Dubonos, and A~A Firsov.
\newblock {Two-dimensional gas of massless Dirac fermions in graphene}.
\newblock {\em Nature}, 438(7065):197--200, November 2005.

\bibitem{Zhang:2005gp}
Yuanbo Zhang, Yan-Wen Tan, Horst~L Stormer, and Philip Kim.
\newblock {Experimental observation of the quantum Hall effect and Berry's
  phase in graphene}.
\newblock {\em Nature}, 438(7065):201--204, November 2005.

\bibitem{Novoselov:2006hu}
K~S Novoselov, E~Mccann, S~V Morozov, V~I Fal'ko, M~I Katsnelson, U~Zeitler,
  D~Jiang, F~Schedin, and A~K Geim.
\newblock {Unconventional quantum Hall effect and Berry's phase of 2$\pi$ in
  bilayer graphene}.
\newblock {\em Nature Physics}, 2(3):177--180, March 2006.

\bibitem{Du:2009ce}
Xu~Du, Ivan Skachko, Fabian Duerr, Adina Luican, and Eva~Y Andrei.
\newblock {Fractional quantum Hall effect and insulating phase of Dirac
  electrons in graphene}.
\newblock {\em Nature}, 462(7270):192--195, November 2009.

\bibitem{Bolotin:2009ko}
Kirill~I Bolotin, Fereshte Ghahari, Michael~D Shulman, Horst~L Stormer, and
  Philip Kim.
\newblock {Observation of the fractional quantum Hall effect in graphene}.
\newblock {\em Nature}, 462(7270):196--199, November 2009.

\bibitem{Zhang:2006hn}
Y~Zhang, Z~Jiang, J~P Small, M~S Purewal, Y~W Tan, M~Fazlollahi, J~D Chudow,
  J~A Jaszczak, H~L Stormer, and P~Kim.
\newblock {Landau-Level Splitting in Graphene in High Magnetic Fields}.
\newblock {\em Physical Review Letters}, 96(13):136806, April 2006.

\bibitem{Laughlin:1983hk}
R~B Laughlin.
\newblock {Anomalous Quantum Hall Effect: An Incompressible Quantum Fluid with
  Fractionally Charged Excitations}.
\newblock {\em Physical Review Letters}, 50(18):1395--1398, May 1983.

\bibitem{Stormer:1999iz}
Horst Stormer.
\newblock {Nobel Lecture: The fractional quantum Hall effect}.
\newblock {\em Reviews of Modern Physics}, 71(4):875--889, July 1999.

\bibitem{Kechedzhi:2012ij}
K~Kechedzhi, E~H Hwang, and S~das Sarma.
\newblock {Gate-tunable quantum transport in double-layer graphene}.
\newblock {\em Physical Review B}, 86(16):165442, October 2012.

\bibitem{Song:2013ji}
Justin Song, Andrey Shytov, and Leonid Levitov.
\newblock {Electron Interactions and Gap Opening in Graphene Superlattices}.
\newblock {\em Physical Review Letters}, 111(26):266801, December 2013.

\bibitem{Bokdam:2014vm}
Menno Bokdam, Taher Amlaki, Geert Brocks, and Paul~J. Kelly.
\newblock {Band gaps in incommensurable graphene on hexagonal boron nitride}.
\newblock {\em arXiv}, page 1401.6207, January 2014.

\bibitem{Hofstadter:1976js}
Douglas Hofstadter.
\newblock {Energy levels and wave functions of Bloch electrons in rational and
  irrational magnetic fields}.
\newblock {\em Physical Review B}, 14(6):2239--2249, September 1976.

\bibitem{Zomer:2014wu}
P.~J. Zomer, M~H~D Guimar{\~a}es, J~C Brant, N.~Tombros, and B.~J. van Wees.
\newblock {Fast pick up technique for high quality heterostructures of bilayer
  graphene and hexagonal boron nitride}.
\newblock {\em arXiv}, page 1403.0399, March 2014.

\bibitem{Geim:2013dq}
A~K Geim and I~V Grigorieva.
\newblock {Van der Waals heterostructures}.
\newblock {\em Nature}, 499(7459):419--425, July 2013.

\bibitem{Georgiou:2012cd}
Thanasis Georgiou, Rashid Jalil, Branson~D Belle, Liam Britnell, Roman~V
  Gorbachev, Sergey~V Morozov, Yong-Jin Kim, Ali Gholinia, Sarah~J Haigh, Oleg
  Makarovsky, Laurence Eaves, Leonid~A Ponomarenko, Andre~K. Geim, Kostya~S
  Novoselov, and Artem Mishchenko.
\newblock {Vertical field-effect transistor based on graphene--WS2
  heterostructures for flexible and transparent electronics}.
\newblock {\em Nature Nanotechnology}, 8(2):100--103, February 2013.

\bibitem{Britnell:2013ca}
Liam Britnell, R~M Ribeiro, A~Eckmann, R~Jalil, B~D Belle, A~Mishchenko, Y~J
  Kim, R~V Gorbachev, T~Georgiou, and S~V Morozov.
\newblock {Strong light-matter interactions in heterostructures of atomically
  thin films}.
\newblock {\em Science (New York, NY)}, 340:1311--1314, June 2013.

\bibitem{Ding:2011eo}
Xuli Ding, Guqiao Ding, Xiaoming Xie, Fuqiang Huang, and Mianheng Jiang.
\newblock {Direct growth of few layer graphene on hexagonal boron nitride by
  chemical vapor deposition}.
\newblock {\em Carbon}, 49(7):2522--2525, June 2011.

\bibitem{Son:2011gk}
Minhyeok Son, Hyunseob Lim, Misun Hong, and Hee~Cheul Choi.
\newblock {Direct growth of graphene pad on exfoliated hexagonal boron nitride
  surface}.
\newblock {\em Nanoscale}, 3(8):3089--3093, 2011.

\bibitem{Tang:2012ju}
Shujie Tang, Guqiao Ding, Xiaoming Xie, Ji~Chen, Chen Wang, Xuli Ding, Fuqiang
  Huang, Wei Lu, and Mianheng Jiang.
\newblock {Nucleation and growth of single crystal graphene on hexagonal boron
  nitride}.
\newblock {\em Carbon}, 50(1):329--331, January 2012.

\bibitem{Shi:2012dg}
Yumeng Shi, Wu~Zhou, Ang-Yu Lu, Wenjing Fang, Yi-Hsien Lee, Allen~Long Hsu,
  Soo~Min Kim, Ki~Kang Kim, Hui~Ying Yang, Lain-Jong Li, Juan-Carlos Idrobo,
  and Jing Kong.
\newblock {van der Waals Epitaxy of MoS 2Layers Using Graphene As Growth
  Templates}.
\newblock {\em Nano Letters}, 12(6):2784--2791, June 2012.

\end{thebibliography}

\end{document}